\documentclass[sigconf, nonacm]{acmart}

\newcommand\vldbdoi{XX.XX/XXX.XX}
\newcommand\vldbpages{XXX-XXX}
\newcommand\vldbvolume{19}
\newcommand\vldbissue{11}
\newcommand\vldbyear{2026}
\newcommand\vldbauthors{\authors}
\newcommand\vldbtitle{\shorttitle} 
\newcommand\vldbavailabilityurl{URL_TO_YOUR_ARTIFACTS}
\newcommand\vldbpagestyle{empty} 
\usepackage{graphicx}
\usepackage{listings}
\usepackage[table]{xcolor}
\usepackage{amsmath,tcolorbox,makecell}
\usepackage{booktabs}
\usepackage{multirow}
\usepackage{array}
\usepackage{pifont}
\usepackage{tikz}
\usepackage{xspace,subcaption,enumitem}
\usetikzlibrary{patterns}
\usetikzlibrary{positioning}
\usetikzlibrary{fit}
\usepackage{makecell}
\usepackage{etoolbox}
\usepackage[framemethod=tikz]{mdframed}
\usepackage{soul}
\usepackage{enumitem}
\usepackage[normalem]{ulem}

\newrobustcmd{\reva}[1]{\textcolor{blue}{{#1}}}
\newrobustcmd{\revb}[1]{\textcolor{darkgreen}{{#1}}}
\newrobustcmd{\revc}[1]{\textcolor{magenta}{{#1}}}
\newrobustcmd{\revm}[1]{\textcolor{darkred}{{#1}}}

\definecolor{codegreen}{rgb}{0,0.6,0}
\definecolor{codegray}{rgb}{0.5,0.5,0.5}
\definecolor{codepurple}{rgb}{0.58,0,0.82}
\definecolor{darkgreen}{rgb}{0,0.5,0}
\definecolor{darkred}{rgb}{0.7, 0.1, 0.12}

\lstdefinestyle{sqlcodestyle}{
    commentstyle=\color{codegreen},
    keywordstyle=\color{codepurple},
    numberstyle=\tiny\color{codegray},
    stringstyle=\color{magenta},
    basicstyle=\ttfamily\footnotesize,
    breakatwhitespace=false,         
    breaklines=true,                 
    captionpos=b,                    
    keepspaces=true,                 
    numbersep=5pt,                  
    showspaces=false,                
    showstringspaces=false,
    showtabs=false,                  
    tabsize=2
}

\lstdefinestyle{inlinesqlcodestyle}{
    commentstyle=\color{codegreen},
    keywordstyle=\color{codepurple},
    numberstyle=\tiny\color{codegray},
    stringstyle=\color{magenta},
    basicstyle=\ttfamily\scriptsize,
    breakatwhitespace=false,         
    breaklines=true,                 
    captionpos=b,                    
    keepspaces=true,                 
    numbersep=5pt,                  
    showspaces=false,                
    showstringspaces=false,
    showtabs=false,                  
    tabsize=2
}

\DeclareMathOperator*{\argmin}{argmin}

\newcommand{\aIns}{D}

\newcommand{\Q}{Q}

\newcommand{\Vset}{\mathcal{V}}
\newcommand{\Qset}{\mathcal{Q}}
\newcommand{\schema}{S}
\newcommand{\V}{\texttt{V}}
\newcommand{\MV}{\texttt{M}}
\newcommand{\R}{\texttt{R}}
\newcommand{\MVset}{\mathcal{M}}
\newcommand{\cost}{\mathsf{cost}}
\newcommand{\size}{\mathsf{size}}

\newcommand{\bagintersection}{{\mathbin{\ooalign{$\cap$\cr\hidewidth\raise0.2ex\hbox{\scalebox{0.6}{$+$}}\hidewidth\cr}}}}

\newcommand{\starrocks}{\mbox{\texttt{STR}}}
\newcommand{\doris}{\mbox{\texttt{DRS}}}

\newcommand{\hive}{\mbox{\texttt{HIV}}}
\newcommand{\calcite}{\mbox{\texttt{CAL}}}

\newcommand{\pg}{\mbox{\texttt{PG}}}
\newcommand{\imdb}{\mbox{IMDB}}
\newcommand{\job}{\mbox{JOB}}
\newcommand{\stats}{\mbox{STATS}}
\newcommand{\scale}{\mbox{SCALE}}
\newcommand{\tpcds}{\mbox{TPC-DS}}
\newcommand{\dsb}{\mbox{DSB}}

\newcommand{\bigsubs}{\mbox{\texttt{BigSubs}}}
\newcommand{\ecse}{\mbox{\texttt{ECSE}}}
\newcommand{\mvpp}{\mbox{\texttt{MVPP}}}
\newcommand{\uniview}{\mbox{\texttt{UniView}}}
\newcommand{\hawc}{\mbox{\texttt{Basic}}}

\newcommand{\gnn}{\mbox{\texttt{GNN}}}
\newcommand{\gnnmv}{\mbox{\texttt{GnnMV}}}
\newcommand{\redshift}{\mbox{\texttt{Sys-A}}}
\newcommand{\celerdata}{\mbox{\texttt{Sys-B}}}
\newcommand{\bigquery}{\mbox{\texttt{Sys-C}}}
\newcommand{\autoview}{\mbox{\texttt{AutoView}}}
\definecolor{GrayRew}{gray}{0.85}
\newcommand{\RCOMMENT}[1]{\noindent
\begin{mdframed}[hidealllines=true,backgroundcolor=GrayRew,skipabove=2mm,skipbelow=2mm,%
    innertopmargin=2mm,%
    innerbottommargin=2mm] #1
\end{mdframed}}

\usetikzlibrary{patterns, 
    fit,
    patterns.meta,
    backgrounds, 
    }

\tikzset{
    fancycellbase/.style={
            rounded corners=2pt,
            inner sep=1pt,
            text height=1.3ex,
            text depth=.0ex,
            text centered,
            font=\scriptsize,
            text=black
    }}
    
\tikzset{
    fancycellhatchbase/.style={
            opacity=0.5,
            rounded corners=2pt,
            fit=(X),
            inner sep=-0.1pt,
    }}

\tikzset{
    fancycellhatch1/.style={
            fancycellhatchbase,
            opacity=0.7,
            pattern={
                Lines[angle=-45,
                distance={3pt}, 
                line width=1pt]},
            pattern color = white,
    }
}

\tikzset{
    fancycellhatch2/.style={
            fancycellhatchbase,
            opacity=0.7,
            pattern={
                Lines[angle=90,
                distance={3pt}, 
                line width=1pt]},
            pattern color = white,
    }
}

\tikzset{
    fancycellhatch3/.style={
            fancycellhatchbase,
            opacity=0.7,
            pattern={
                Dots[angle=90,
                distance={3pt}, 
                radius=0.8pt]},
            pattern color = white,
    }
}

\tikzset{
    fancycellhatch4/.style={
            fancycellhatchbase,
            opacity=0.7,
            pattern={
                Hatch[angle=45,
                distance={6pt}, 
                line width=0.7pt]},
            pattern color = white,
    }
}

\newcommand{\fancycellC}[5]{%
  \begingroup
    \definecolor{__tmpfill}{HTML}{#2}%
    \begin{tikzpicture}[baseline=(X.base)]
        \node[fancycellbase,
            minimum width=#1,
        ] (X) {#3};
        \begin{scope}[on background layer]
            \node[fancycellhatchbase, fill=__tmpfill!70]{};
            \node[#4] {}; 
        \end{scope}
    \end{tikzpicture}%
  \endgroup
}

\newcommand{\fancycellGreen}[1]{\makebox[0pt][c]{\fancycellC{2.8em}{1e7e34}{#1}{}{ $\checkmark$ }}}
\newcommand{\fancycellYellow}[1]{\fancycellC{2.8em}{b8860b}{#1}{}{ $\blacktriangle$ }}

\newcommand{\fancyCellRed}[1]{\fancycellC{0em}{c81b4a}{ \xmark }{fancycellhatch4}{}}

\tikzset{
    fancycellbaseNormalTextSize/.style={
            rounded corners=2pt,
            inner sep=1pt,
            text height=1.3ex,
            text depth=.0ex,
            text centered,
            font=\normalsize,
            text=black
    }}
    
\newcommand{\fancycellCNormal}[5]{%
  \begingroup
    \definecolor{__tmpfill}{HTML}{#2}%
    \begin{tikzpicture}[baseline=(X.base)]
        \node[fancycellbaseNormalTextSize,
            minimum width=#1,
        ] (X) {#3};
        \begin{scope}[on background layer]
            \node[fancycellhatchbase, fill=__tmpfill!70]{};
            \node[#4] {}; 
        \end{scope}
    \end{tikzpicture}%
  \endgroup
}

\newcommand{\fancycellGreenNormal}[1]{\fancycellCNormal{2.8em}{1e7e34}{#1}{}{ $\checkmark$ }}
\newcommand{\fancycellYellowNormal}[1]{\fancycellCNormal{2.8em}{b8860b}{#1}{}{ $\blacktriangle$ }}
\newcolumntype{C}[1]{>{\centering\arraybackslash}p{#1}}
\theoremstyle{definition}

\theoremstyle{remark}

\begin{document}

\title{Benchmarking the Full Pipeline of Materialized-View-Based Query Rewriting}

\author{Xinjie Hu}
\affiliation{%
  \institution{Simon Fraser University}
  \city{Burnaby}
  \country{Canada}
}
\email{xha102@sfu.ca}

\author{Zhengjie Miao}
\affiliation{%
  \institution{Simon Fraser University}
  \city{Burnaby}
  \country{Canada}}
\email{zhengjie@sfu.ca}

\begin{abstract}
Materialized views (MVs) accelerate OLAP and data-warehouse workloads by precomputing reusable subexpressions, but practical MV-based query acceleration is a multi-stage pipeline: candidate enumeration, view selection under storage budgets, and query rewriting inside the optimizer. Existing evaluations typically study only parts of this pipeline and within a single system, leaving end-to-end trade-offs and cross-system behavior unclear.

In this paper, we benchmark MV-based query rewriting by jointly evaluating enumeration, selection, and rewriting with a modular evaluation framework and by using controlled ablations. We also introduce a cross-engine protocol allowing us to compare systems that expose only execution plans by contrasting native optimizer-level rewriting with portable SQL rewriting baselines when available. Across representative academic methods and modern open-source and commercial systems, we find strong interaction effects across stages and large variability in MV usage and realized savings.
We identify recurring failure modes that explain performance regressions after rewriting.
Our results highlight which pipeline stages most often limit performance and provide evidence to guide future MV enumeration, selection, and rewriting designs.
\end{abstract}

\maketitle

\pagestyle{\vldbpagestyle}
\begingroup\small\noindent\raggedright\textbf{PVLDB Reference Format:}\\
\vldbauthors. \vldbtitle. PVLDB, \vldbvolume(\vldbissue): \vldbpages, \vldbyear.\\
\href{https://doi.org/\vldbdoi}{doi:\vldbdoi}
\endgroup
\begingroup
\renewcommand\thefootnote{}\footnote{\noindent
This work is licensed under the Creative Commons BY-NC-ND 4.0 International License. Visit \url{https://creativecommons.org/licenses/by-nc-nd/4.0/} to view a copy of this license. For any use beyond those covered by this license, obtain permission by emailing \href{mailto:info@vldb.org}{info@vldb.org}. Copyright is held by the owner/author(s). Publication rights licensed to the VLDB Endowment. \\
\raggedright Proceedings of the VLDB Endowment, Vol. \vldbvolume, No. \vldbissue\ %
ISSN 2150-8097. \\
\href{https://doi.org/\vldbdoi}{doi:\vldbdoi} \\
}\addtocounter{footnote}{-1}\endgroup

\ifdefempty{\vldbavailabilityurl}{}{
\vspace{.3cm}
\begingroup\small\noindent\raggedright\textbf{PVLDB Artifact Availability:}\\
The source code, data, and/or other artifacts have been made available at \url{https://github.com/edx-h/Benchmarking-MV-Based-Rewriting}.
\endgroup
}

\section{Introduction}
\label{sec:introduction}

As analytical workloads grow in scale and complexity, many queries repeatedly execute expensive subexpressions (e.g., joins and aggregations) over large base tables, leading to high latency and substantial redundant computation. Materialized views (MVs) offer a principled way to eliminate this redundancy: by precomputing and storing reusable subexpressions, a DBMS can rewrite incoming queries to reuse materialized results, replacing repeated computation with a (typically cheaper) view scan. 
This motivates MV-based query rewriting, which is now supported in an increasing number of database systems (automatic MV rewrite support: Oracle 2009, Snowflake 2018, RedShift 2020, BigQuery 2021, StarRocks 2023, Doris 2024~\cite{oracle,snowflake,redshift,bigquery,starrocks,doris}.)

Conceptually, MV reuse is straightforward, whereas realizing MV-based rewriting in practice requires a multi-stage pipeline whose components depend on one another: given a workload, an \emph{enumerator} generates candidate views; a \emph{selector} chooses a subset under a storage budget to maximize expected runtime savings; and a \emph{rewriter} decides whether and how to rewrite each query to use the selected views. In practice, production systems differ in what they expose: some expose intermediate outputs such as selected views, while others do not; some expose rewritten SQL, while others perform rewriting internally and only reveal the final execution plan. This limited observability makes it difficult to compare systems --- especially rewriting behavior across engines --- and motivates our cross-platform evaluation protocol.

Existing research targets the pipeline asymmetrically: enumeration~\cite{ho1997range,vldb_2000_automated_mv_selection_microsoft,edbt_2009_chaves_sap} and selection~\cite{vldb_2018_bigsubs_microsoft,arxiv_2019_dqm_chicago,icde_2021_autoview_thu,vldb_1997_mvpp_unsw} have received dedicated empirical study and benchmark-style cross-method comparison, whereas the rewriting stage has been studied algorithmically~\cite{dexa_1998_Ligoudistianos_athens,vldb_1997_Theodoratos_athens,caise_1999_Theodoratos_athens,data_know_2001_Theodoratos_athens,er_1998_Theodoratos_athens} but to our knowledge, no comprehensive benchmark study compares industrial rewriters across engines under a common protocol. This leaves practitioners without empirical guidance on the most engine-dependent stage; combined with the absence of an end-to-end evaluation across all three stages under a unified setting, it makes pipeline behavior and practical trade-offs difficult to assess. To fill this gap, we build a unified evaluation protocol that covers all three stages ---  enumeration, selection, and rewriting --- and introduce a cross-platform protocol to support both pluggable components and systems with limited observability. In this paper, we experimentally investigate the full MV-based rewriting pipeline to answer the following research questions:

\noindent\textbf{RQ1: Is MV-based query rewriting ``solved'' in practice?}
Do end-to-end pipelines deliver \emph{consistent} workload-level savings across workloads and budgets? How often do rewrites regress, and how do \emph{industrial end-to-end solutions} compare with \emph{modular baselines}?

\noindent\textbf{RQ2: How does MV-based query rewriting vary across engines under limited observability?}
When engines expose only plan-level output rather than rewritten SQL, how can we compare their MV exploitation fairly? Under a common MV set, how does native optimizer-level rewriting compare with a portable SQL rewrite baseline executed on the same engine?

\noindent\textbf{RQ3: Which stage is the bottleneck for end-to-end performance, and why?}
When is the limiting factor candidate generation, budget-constrained selection, or the rewriter's ability to exploit selected views? What workload and candidate-space characteristics explain these bottlenecks, and what failure modes recur?

\smallskip

\smallskip

\noindent\textbf{Contributions.}
We provide a systematic evaluation of MV-based query rewriting methods, covering both end-to-end pipelines and individual components. Our contributions are:

\begin{enumerate}[leftmargin=0.5cm]
\item \textbf{End-to-end empirical evaluation across all three stages.}
We develop a modular framework that decomposes MV-based query rewriting into enumeration, selection, and rewriting, enabling systematic comparison and controlled ablations (fix two stages, vary the third). To our knowledge this is the first end-to-end empirical study that jointly evaluates representative methods across all stages under a unified protocol (Section~\ref{sec:architecture}).

\item \textbf{Cross-engine protocol and a systematic comparison of industrial and open-source rewriters.} 
We study a broad set of open-source and commercial systems and propose a cross-engine evaluation protocol that incorporates systems with limited observability (those exposing only plans rather than rewritten SQL) by comparing rewriting on a common execution engine. Using it, we provide the first benchmark comparison of industrial rewriters --- \hive, \calcite, \doris, \starrocks, \redshift, \celerdata, and \bigquery{} --- under a common protocol with shared workloads and view sets, on rewrite success rate, runtime impact, and failure-mode signatures (Sections~\ref{sec:architecture}, \ref{sec:setup}, \ref{sec:exp-rewriter}, \ref{sec:robustness}).

\item \textbf{Reusable implementation and metrics.} We implement and integrate representative baselines across stages and systems behind a unified interface for plugging in additional methods and engines, and curate stage-wise and end-to-end metrics (e.g., workload time saving, per-view query coverage, regression behavior) under a standardized protocol (Sections~\ref{sec:enumeration}--\ref{sec:setup}).
\end{enumerate}

\noindent\textbf{Summary of findings.}
Our results show that isolated stage-wise evaluations are insufficient for understanding MV-based rewriting pipelines. Pipeline outcomes depend on how candidate generation, selection under budget, and rewriting interact. We summarize three quantitative, workload-conditioned findings.

\begin{itemize}[leftmargin=0.4cm]

\item \textbf{Single-stage rankings can be locally valid but globally misleading.}
On PostgreSQL/\job{}/\hive~ rewriter at 1,GB, fixing \bigsubs{} makes \ecse{} outperform \hawc{} by 43.85\,pp, while fixing \hawc{}\ makes \gnnmv{} outperform \bigsubs{} by 47.53\,pp. Yet both \ecse{}--\bigsubs{} and \hawc{}--\gnnmv{} are high-performing pipelines (Figure~\ref{fig:pg_detailed_result}). Thus, enumerators and selectors cannot be ranked independently: different candidate spaces interact with different selectors (\textsection~\ref{sec:stage_wise_analysis}).

\item \textbf{The dominant bottleneck changes with workload and pipeline context.}
Enumeration limits performance when high-coverage views are missing (Figure~\ref{fig:pg_detailed_result}, \textsection \ref{sec:exp-enumerator}); selection dominates when such views must be ranked under tight budgets (Table~\ref{tab:exp-selector-filter-quality} in appendix); and rewriting dominates when selected views are not exploited by the engine (Table~\ref{tab:exp-rewrite-success}, \textsection\ref{sec:exp-rewriter}). These shifts explain why single-stage evaluation under one fixed context misses important failure modes.

\item \textbf{Modular pipelines often beat commercial integrated systems, but not uniformly.}
On \job{}, the best modular pipelines outperform commercial auto-selection + native rewriting by roughly 40\,pp or more (Figure~\ref{fig:exp-cross-engine}); on \tpcds{}, the gap narrows or reverses when the integrated enumeration/selection in the commercial system produces more compact, useful views (through column pruning) that fit into the storage budget (Section~\ref{sec:exp-case}).

\end{itemize}

\noindent\textbf{Outline.} Section~\ref{sec:prelim} defines the MV-based rewriting pipeline. Section~\ref{sec:architecture} presents our modular evaluation framework and cross-engine protocol. Sections~\ref{sec:enumeration}--\ref{sec:rewriter} describe the evaluated enumerators, selectors, and rewriters. Section~\ref{sec:setup} gives experimental setup and metrics. Section~\ref{sec:exp-endtoend} shows end-to-end and cross-engine results, establishing why isolated stage rankings are insufficient. Section~\ref{sec:stage_wise_analysis} then diagnoses the mechanisms behind these results through stage-wise and robustness analyses. Section~\ref{sec:exp-case} provides representative case studies, and Section~\ref{sec:conclusion} concludes with guidance and future directions.
\section{Preliminaries and Background}
\label{sec:prelim}
\begin{figure*}
    \centering
    \includegraphics[width=0.92\linewidth]{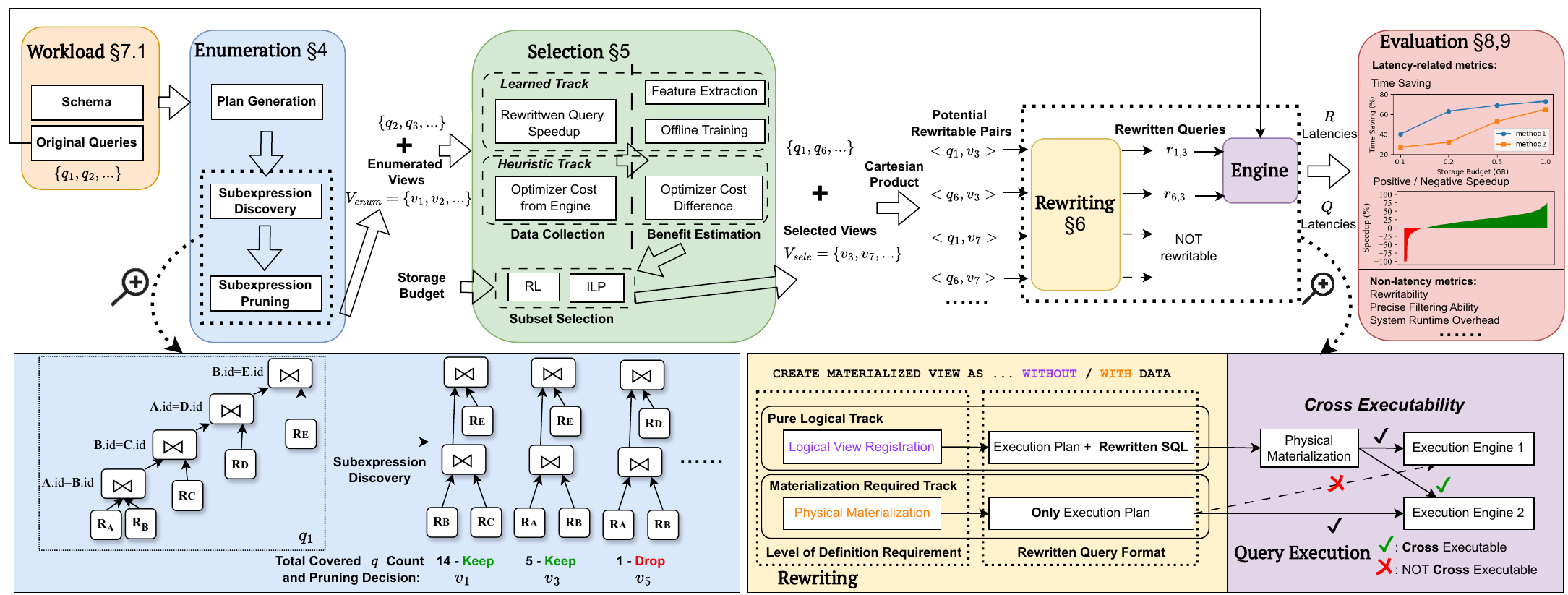}
    \vspace{-3mm}
    \caption{The modular evaluation framework for MV-based query rewriting pipeline. Blocks at the bottom show the details of the components of the same color in the framework. Numbers denote the corresponding section.}
    \label{fig:architecture}
\end{figure*}

Given a set of queries $\Qset = \{\Q_1, …, \Q_m\}$, a database schema $\schema$, and a database instance $\aIns$ of $\schema$, the goal of MV-based query rewriting is to accelerate the execution of queries in $\Qset$ using materialized views, which is further decomposed into three subtasks.

\noindent\textbf{Query Rewrite} Given a query $\Q$, a database instance $\aIns$ of $\schema$, a set of views $\Vset = \{\V_1, \V_2, …, \V_n\}$, views in $\Vset$  are instantiated as materialized views $\MVset = \{\MV_1 = \V_1(\aIns), \MV_2 = \V_2(\aIns), \ldots,\MV_n = \V_n(\aIns)\}$, the query rewrite problem is to find a rewritten query $\R$ of $\Q$ with the minimum $\cost(\R | \aIns, \MVset)$, where $\cost(\R | \aIns, \MVset)$  is the evaluation cost of $\R$ over the database instance $\aIns$ with views materialized.

\noindent\textbf{View Selection} Given $\Qset = \{\Q_1, \ldots, \Q_m\}$ and a set of materialized views $\MVset = \{\MV_1 = \V_1(\aIns), \MV_2 = \V_2(\aIns), \ldots,\MV_n = \V_n(\aIns)\}$, for any non-empty subset $\Vset' \subseteq \Vset$, the view selection problem is to find a subset of MVs $\MVset^* \subseteq \MVset$ to minimize the cost of evaluating queries in $\Qset$ within a storage budget constraint:
\[\MVset^* = \argmin_{\MVset' \subseteq \MVset, \size(\MVset') \leq B }  \sum_{i=1}^{m} \cost(\R_i | \aIns, \MVset')\]
where $\R_i$ is a rewrite of query $\Q_i$, and $\cost(\R_i | \aIns,\MVset')$ denotes the execution cost of $\R_i$ on $\aIns$ when the MVs in $\mathcal{M}'$ are available, and $\size(\mathcal{M}')$ is the storage cost under budget $B$.

\noindent\textbf{Candidate View Enumeration.} Given $\Qset = \{\Q_1, \ldots, \Q_m\}$ and $\schema$,  candidate view enumeration (enumeration for short) is a procedure that outputs a finite set of view definitions $\Vset=\{\V_1,\ldots,\V_n\}$ to be considered by the selector.
In practice, $\Vset$ is often derived from query subexpressions (e.g., joins/aggregations) and may apply pruning rules to control the candidate set size.

\section{Modular Evaluation Framework}
\label{sec:architecture}

We design a modular evaluation framework (Figure~\ref{fig:architecture}). Given a query workload, an \emph{enumerator} generates a set of candidate views. A \emph{selector} chooses a subset of candidates under a storage budget. Finally, a \emph{rewriter} rewrites workload queries by substituting view scans for matching subexpressions, returning either rewritten SQL (when exposed) or an execution plan.

\noindent\textbf{Pluggability.}
We call a pipeline component \emph{pluggable} if it exposes its intermediate outputs through a common interface (e.g., candidate views, selected views, or explicit rewrites), allowing us to substitute alternative implementations without modifying the rest of the pipeline. In practice, pluggability differs across stages: most enumerators and recommenders are pluggable with exposing view definitions and/or selected views, whereas most rewriting is performed only in \emph{plan-transparent} way, i.e., merely observable through the final \texttt{EXPLAIN} plan rather than an explicit rewritten SQL query. We determine rewrite success by checking whether the returned a rewritten SQL or the plan references a materialized view.

\noindent\textbf{Cross-platform evaluation.}
For fully pluggable pipelines, our framework enables evaluation over combinations of methods, i.e., $\{Enumerators\}\times\{Selectors\}\times\{Rewriters\}$. For pipelines with non-pluggable components (most commonly plan-transparent rewriters), we use a cross-platform protocol. At a high level, we evaluate optimizer-driven rewriting on the target engine through plan inspection and runtime, and additionally execute portable rewritten SQL produced by a SQL-transparent rewriter on the same engine. Details of the cross-platform protocol are provided in Section~\ref{sec:setup}.
\section{Candidate View Enumeration}
\label{sec:enumeration}

\subsection{View Enumeration Methods}

\begin{figure}[htbp]
    \centering
    \includegraphics[width=\linewidth]{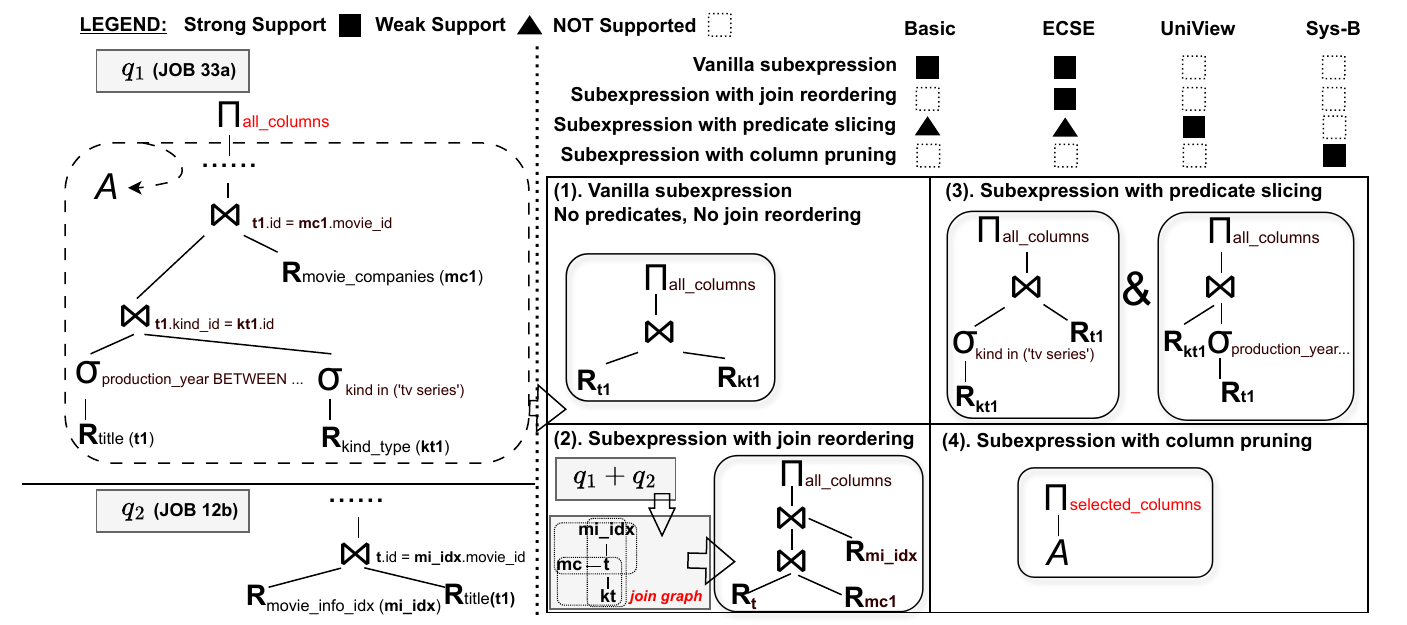}
    \vspace{-3mm}
    \caption{Behavior Comparison of Enumerators. We take 33a, 12b from \job~ as two original queries for illustration.}
\label{fig:enumerator_different_behaviour}
\end{figure}

Candidate view enumeration generates candidate view definitions from a workload. Prior approaches can be organized by how they define and prune the candidate space. Classical approaches enumerate structured search spaces such as data-cube lattices or AND/OR view graphs~\cite{sigmod_1996_Harinarayan_stanford,ho1997range,gupta1997selection,Gupta2005Selection}. Workload-subexpression approaches~\cite{vldb_2000_automated_mv_selection_microsoft,Zhou2007Efficient,giannikis2014shared,vldb_2018_bigsubs_microsoft,Jindal2018Selecting,Jindal2018Computation,sigmod_2001_optimizing_using_mv_microsoft} extract candidates from query plans or shared subplans, often with pruning based on frequency, cost, or containment relationships. Join-graph approaches further expand the candidate space by deriving workload-level join structures beyond a single optimizer plan~\cite{vldb_2020_Rafi_oracle}. MVPP-style approaches merge workload queries into a shared DAG and select reusable intermediate nodes as candidates~\cite{vldb_1997_mvpp_unsw,icde_2021_autoview_thu,soft_computing_2003_Horng,vldb_2024_uniview_zju}. Finally, production systems may couple enumeration with selection and expose only the resulting recommended views. The appendix provides a more detailed survey; here we focus on the representative enumerators used in our evaluation (Table~\ref{tab:enum_mapping}).

\subsection{Evaluated Methods and Implementation}

We evaluate the four representative enumerators in Table~\ref{tab:enum_mapping} (three academic methods plus the commercial \celerdata{}). The evaluated enumerators differ along three axes: join-structure expansion, predicate slicing, and column projection (Figure~\ref{fig:enumerator_different_behaviour}); we describe each in two to three sentences below and leave the detailed per-query behavior to the figure.

\begin{table}[h]\centering
\caption{Evaluated enumerators and their taxonomy.}\label{tab:enum_mapping}
\vspace{-3mm}
\scriptsize
\begin{tabular}{p{0.8cm}p{1.9cm}p{2.0cm}p{2.0cm}}\toprule
\textbf{Method} & \textbf{Category} & \textbf{Key behavior} & \textbf{Why included} \\\midrule
\hawc & Plan-subtree & Bottom-up plan subexpressions & Classical heuristic baseline \\
\ecse & Join-graph / workload & Explores join-order variants & Tests structural expansion \\
\uniview & MVPP  & Predicate-rich merged DAG views & Recent open-source method \\
\celerdata & Industrial integrated & Coupled enum./selection & Production system behavior \\
\bottomrule
\end{tabular}
\end{table}

\noindent\textbf{\hawc}~\cite{sigmod_1996_Harinarayan_stanford, dolap_1999_HIDETOSHI_michigan} is a classical heuristic enumerator that builds candidates bottom-up from the query plan, adding one join relation per step, so every candidate is a subtree of the chosen plan. It preserves all columns and slices predicates only conservatively (after all relations are joined) to limit candidate explosion, and consequently cannot discover join orders absent from that plan.

\noindent\textbf{\ecse}~\cite{vldb_2020_Rafi_oracle} expands the candidate space by discovering join patterns shared across queries but absent from any single plan~\cite{sigmod_2000_roy_iit,sigmod_2001_hoshi_iit,arxiv_2002_Valluri_iit,arxiv_2019_dqm_chicago,sigmod_2001_optimizing_using_mv_microsoft,vldb_2020_Rafi_oracle}: it maintains a workload-level join graph and takes the transitive closure over equality predicates to synthesize new join structures. Its predicate and projection handling follows \hawc{}.

\noindent\textbf{\uniview}~\cite{vldb_2024_uniview_zju}, the only open-source enumerator we found, represents \mvpp-style methods~\cite{vldb_1997_mvpp_unsw,icde_2021_autoview_thu,soft_computing_2003_Horng}: it merges the workload's queries into a single DAG whose non-leaf nodes become candidates. Its implementation slices predicates so that every candidate carries at least one predicate (predicate-free views are pruned).

\noindent\textbf{\celerdata{}} couples enumeration with selection inside a commercial engine, generating candidates internally; we characterize its aggressive column pruning in the case study in \textsection\ref{sec:exp-case}.

\section{View Selection}\label{sec:recommendation}

\subsection{View Selection Methods}

View selection chooses a subset of candidate MVs under a storage budget. Prior work differs mainly in how it estimates view benefit and how it searches the budget-constrained subset space. Classical methods rely on optimizer cost estimates and select views using greedy, dynamic-programming, or integer-programming algorithms over structured candidate spaces such as cube lattices or AND/OR view graphs~\cite{sigmod_1996_Harinarayan_stanford,gupta1997selection,Gupta2005Selection,data_eng_conference_2014_perez_rice,vldb_2018_bigsubs_microsoft}. Other approaches use metaheuristics such as genetic search, represented by \mvpp{}~\cite{vldb_1997_mvpp_unsw}. More recent learned methods, including DQM, AutoView, and UniView, learn query/view representations and cost or benefit models, sometimes combining with reinforcement learning~\cite{arxiv_2019_dqm_chicago,icde_2021_autoview_thu,vldb_2024_uniview_zju}. Commercial systems such as \celerdata{} further couple enumeration and selection into an integrated black-box recommender. Table~\ref{tab:selector_mapping} maps these categories to the selectors evaluated in this paper.

\subsection{Evaluated Methods}
MV selection can be divided into two sub-stages: estimating the benefit of each candidate view and selecting a subset under the storage budget. We evaluate four representative selectors, including one commercial integrated system. Table~\ref{tab:selector_mapping} summarizes their benefit model, subset-search strategy, and role in our evaluation.

\begin{table}[h]
\centering
\caption{Evaluated selectors.}
\label{tab:selector_mapping}
\scriptsize
\vspace{-3mm}
\begin{tabular}{p{0.75cm}p{1.7cm}p{1.45cm}p{3.0cm}}
\toprule
\textbf{Method} & \textbf{Benefit model} & \textbf{Subset search} & \textbf{Why included} \\
\midrule
\bigsubs{} & Optimizer-cost proxy & ILP & Classical cost-based method \\
\gnnmv{} & Learned utility & ILP & Learned ranking with budgeted selection \\
\uniview{} & Learned cost model & RL/DQN & Recent open-source learned selector \\
\celerdata{} & System-internal & System-internal & Representative industrial selector \\
\bottomrule
\end{tabular}
\end{table}

\noindent\textbf{\bigsubs}~\cite{vldb_2018_bigsubs_microsoft} formulates selection as an ILP that maximizes the total cost-proxy utility of selected views (per-query reduction $\cost(q)-\cost(q\mid v)$) under the budget, with overlap constraints discouraging redundant views (full formulation in appendix).

\noindent\textbf{\gnnmv}~\cite{gnn} replaces the cost proxy with a graph neural network (GNN) that predicts per-view utility from the structure of queries, candidate views, and base tables, then selects under budget with ILP. The model is trained offline using features extracted from query workloads, including query structure statistics and estimated costs.

\noindent\textbf{\uniview}~\cite{vldb_2024_uniview_zju} uses a deep neural network (DNN) to estimate post-rewrite execution cost and a Deep Q-Network (with reinforcement learning) to choose the view set.

\noindent\textbf{\celerdata{}} couples enumeration and selection in the backend and exposes only the selected views, with an undisclosed internal benefit model. 
By parsing historical query logs, it identifies recurring subplans to extract structured SPJG patterns.

\noindent\underline{Implementation.}
Most prior MV-selection systems are not publicly available. We therefore implement \bigsubs{} using its ILP formulation and implement \gnnmv{} following the model and training procedure described in its paper. For \uniview{}, we use its public implementation. The learning-based selectors are trained separately for each dataset using the corresponding training workload. For \celerdata{}, we use the system-provided recommender and treat its internal selection logic as a black box.
\section{Rewriter}\label{sec:rewriter}
\subsection{Rewriting Queries Using Views}

Query rewriting using views has a long history in data warehouses and data integration systems~\cite{vldb_2001_Halevy_washington,sigmod_1996_Harinarayan_stanford,vldb_1996_levy_at&t,IJCAI_1999_eric}. Classical work studies rewriting through query containment and equivalence, first for conjunctive queries and later for richer SQL fragments with aggregation, comparison predicates, and set operations~\cite{stoc_1977_Chandra_ibm,acm_1981_sagiv_princeton,pods_1999_Grumbach_inria,vldb_1995_gupta_ibm,pods_1995_levy_at&t,Cohen2006Rewriting}. These extensions substantially increase the complexity of deciding whether and how a query can be rewritten using available views~\cite{Abiteboul1995Foundations,vldb_2001_Halevy_washington}.
Modern DBMSs typically implement MV rewriting inside the query optimizer rather than as a standalone rule system~\cite{vldb_1998_bello_oracle,icde_1995_Chaudhuri_hplab,vldb_1996_Tsatalos_ibm}. Under this architecture, view substitutions are considered as part of plan search and interact with normalization, rule matching, and cost estimation. Despite this mature algorithmic and system literature, to our knowledge no prior study systematically benchmarks practical MV rewriters across engines under a common protocol. This motivates our evaluation of both SQL-transparent and plan-transparent rewriters under a common cross-engine protocol and the cross-engine analysis in Section~\ref{sec:exp-rewriter}.

\subsection{Evaluated methods}

Most academic rewriting algorithms are either conceptual or not publicly available as robust SQL-level tools. The only open-source academic system we found with an implemented rewriting component is \uniview{}, but in our preliminary validation it produced a non-negligible number of non-equivalent rewrites, so we exclude it from the rewriter comparison and focus on systems whose rewrites are validated through production query optimizers. We evaluate seven practical rewriters spanning SQL-transparent and plan-transparent systems.

\noindent\textbf{\calcite} (vanilla Calcite~\cite{calcite}) applies built-in \texttt{SubstitutionVisitor} class and default view-related rules over relational-algebra plans, with no engine-specific tuning.

\noindent\textbf{\hive} (Apache Hive~\cite{hive}) embeds Calcite-based view rewriting directly inside its optimizer and, with \calcite, is one of the two rewriters that emit portable rewritten SQL.

\noindent\textbf{\starrocks} (StarRocks~\cite{starrocks}) and \textbf{\doris} (Doris~\cite{doris}) are open-source MPP engines with native optimizer-level MV rewriting; \starrocks{} documents that its design is inspired by prior work on optimizing queries using materialized views~\cite{sigmod_2001_optimizing_using_mv_microsoft}\footnote{\url{https://github.com/StarRocks/starrocks/issues/11386}}.

\noindent \textbf{\redshift}, \textbf{\celerdata}, and \textbf{\bigquery} are commercial cloud systems that provide an automatic query rewrite feature inside the optimizer; their optimizer-level MV rewriting is exposed through execution plans rather than rewritten SQL. We treat them as plan-transparent rewriters in our cross-engine protocol.

\noindent \textbf{Cross-engine executability.} 
Only \calcite{} and \hive{} expose rewritten SQL; these rewrites can be executed on other engines after dialect translation when the translated SQL preserves semantics (we used \texttt{sqlglot} in our implementation). The remaining rewriters are plan-transparent: they expose only the final execution plan, and we infer rewrite success when the plan references a materialized view. This observability gap motivates our cross-engine protocol (Section~\ref{sec:setup}): for each plan-transparent engine, we compare its native optimizer-level rewriting against a portable \hive{} SQL rewrite executed on the same engine under the same materialized view set.
\section{Experimental Setup}
\label{sec:setup}

\subsection{Datasets and Workloads.} 
We perform the experiments with four workloads, namely \job~ \cite{vldb_2015_imdb_job_tum}, \scale~ \cite{scale}, \stats~ \cite{vldb_2021_cardinality_benchmark_(stats)_alibaba,sigmod_2025_athena_sustech} and \tpcds~ (SF=10) \cite{tpcds}, with only \job~ and \stats~ sharing the same schema and dataset \imdb~ \cite{vldb_2015_imdb_job_tum}. We choose them because they are widely used in the query optimization and view-based query rewriting literature \cite{arxiv_2019_dqm_chicago,gnn,icde_2021_autoview_thu,vldb_2024_uniview_zju}. Additionally, these workloads cover different difficulties due to their significant diversities measured by join count (Table \ref{tab:workload_query_cnt_join_cnt} in appendix) and operator count in selection predicate (Table \ref{tab:workload_selection_predicate_operator-stat} in appendix). \job~ is the most complex one due to the greatest join count and more logical connectives, null judgement, string operations; \stats~ and \tpcds~ have comparable join counts while the former involves more binary comparisons and type conversion, and the latter presents more arithmetic and range computations; \scale~ is the simplest one with the lowest join count and fewest operator occurrences.

\subsection{Platform and Implementation}
\noindent\textbf{Execution Engine and Cross Executability.} 
We evaluate the following seven engines: PostgreSQL, Hive, Doris, StarRocks, \redshift, \celerdata~ and \bigquery. For engines whose rewriting is plan-transparent (Doris, StarRocks, \redshift, \celerdata, \bigquery), we compare two rewriters on the same engine: (i) the engine’s native optimizer-level rewriting (measured via \texttt{EXPLAIN} and execution latency), and (ii) executing portable rewritten SQL produced by Hive. On PostgreSQL, we additionally include Calcite as a second SQL-transparent baseline. This instantiates the cross-platform protocol defined in Section~\ref{sec:architecture}.

\noindent\textbf{Training for learned selectors.} 
To support learned view selection baselines (\uniview~ and \gnnmv) and ensure a fair comparison, we split the original workload queries into training/validation/test sets with a 3:1:6 ratio, following prior work~\cite{gnn}. All methods are evaluated on the same test set. 
To construct training instances for learned selectors, we generate query--view pairs and label their utility as described in Section~\ref{sec:recommendation}: i) use \hive~ to attempt a rewrite of a query with a candidate view; ii) execute both the original and rewritten queries on PostgreSQL and record their latencies; and iii) use the latency reduction as the utility label for the pair.

\noindent \textbf{Storage Budget.} The storage budget bounds the \textbf{total on-disk size of all selected materialized views} --- nor query-result caches, nor main-memory buffers. We measure view size using PostgreSQL's \texttt{pg\_total\_relation\_size}, which includes the table heap and any indexes the selector would create. The primary budget is 1\,GB; for sensitivity analysis we additionally sweep 0.1, 0.2, 0.5\,GB (the budgets are comparable to those used in prior work on MV selection~\cite{gnn,icde_2021_autoview_thu,vldb_2024_uniview_zju} for the datasets). Given the underlying database sizes (\imdb~ 8.31\,GB, used by \job/\scale; \tpcds~ 21.96\,GB), these budgets correspond to roughly 1.2--12\% of \imdb~ and 0.5--4.5\% of \tpcds, spanning tight ($<$2\% of the base data) through moderate ($\sim$10\%) regimes. This range overlaps the budgets used by \bigsubs~\cite{vldb_2018_bigsubs_microsoft} (typically $\leq$2\%) and \uniview~\cite{vldb_2024_uniview_zju} (up to $\sim$10\%), enabling direct comparison with prior work. The fact that 1\,GB is moderate on \imdb~ but tight on \tpcds~ is itself a finding (see the column-pruning case study in Section~\ref{sec:exp-case}).

\noindent\textbf{Hardware and Software.}
We run all non-commercial systems on an AWS EC2 instance. The instance model is c5a.16xlarge (64 vCPUs, 128 GB RAM, gp3 EBS and Ubuntu 24.04 OS). For \redshift~ and \celerdata, the vendor-provided service stacks are also deployed on AWS.

\subsection{Metrics}

\noindent\textbf{End-to-end metric.}
Since our evaluation considers a large number of enumerator--selector--rewriter combinations, we use a single end-to-end metric, \emph{relative workload time saving}, to summarize the overall effectiveness of each combination. If a rewritten query is slower than its original counterpart, we treat it as \emph{no rewrite} (i.e., we use the original query latency when aggregating workload time saving). All other metrics (including speedup vs regression) are reported at the stage level (enumeration, selection, and rewriting) to diagnose the sources of performance differences.

Unless otherwise stated, we evaluate each component using the best-performing combination of the other two components on each dataset within a fixed search space, e.g., when comparing rewriters, the input to each rewriter is the same set of views selected from a fixed combination of enumerator and selector

\noindent\textbf{Enumeration.} We evaluate candidate enumerators using metrics for both scalability and candidate-set quality:
(i) \emph{Candidate set size} (number of generated view candidates);
(ii) \emph{Materialized view size distribution} (which affect downstream selection under budgets);
(iii) \emph{Candidate complexity}, including average join size (number of joined tables), the diversity of join orders, and the number/type of predicate operators in \texttt{WHERE};
(iv) \emph{Per-view query coverage}, measured from the rewrite-relationship graph between views and queries: the distribution of how many queries a view can rewrite.

\noindent\textbf{View selection.}
We evaluate selection methods under the same candidate set across different storage budgets, focusing on both effectiveness and overhead:
(i) \emph{Selection overhead}, including solver runtime for ILP-based methods and training/inference costs for learned methods;
(ii) \emph{Filtering quality} (Section \ref{sec:precise_filtering} in appendix), i.e., how effectively a selector discards useless candidates. We set selection, rewriting and acceleration as three checkpoints and compare recommenders' behaviors.

\noindent\textbf{Query rewriting.}
For rewriters, we report:
(i) \emph{Rewrite success rate} (fraction of queries for which a view is used, as indicated by the plan from \texttt{EXPLAIN} or by rewritten SQL when available);
(ii) \emph{Runtime impact}, including overall speedup distribution and the fraction of queries that improve versus regress.

\section{End-to-end and Cross-engine Performance (RQ1, RQ2)}

\label{sec:exp-endtoend}
We therefore report end-to-end results across all three stages under two complementary settings.

\noindent\textbf{(i) PostgreSQL end-to-end.} We evaluate the full modular pipeline on PostgreSQL, where candidate views, selected views, and rewritten SQL are observable and portable. This enables controlled comparisons across enumerators, selectors, and SQL-transparent rewriters by executing all rewritten queries on a common engine.

\noindent\textbf{(ii) Cross-engine end-to-end.} To incorporate systems that offer plan-transparent rewriting (i.e., exposed only through \texttt{EXPLAIN}), we execute the original workload on each engine. We then compare each engine's native optimizer-level rewriting against a portable SQL-rewrite baseline produced by Hive and execute the rewritten queries on the same engine under the same materialized-view set.

\begin{figure*}
    \centering
    \includegraphics[width=0.92\linewidth]{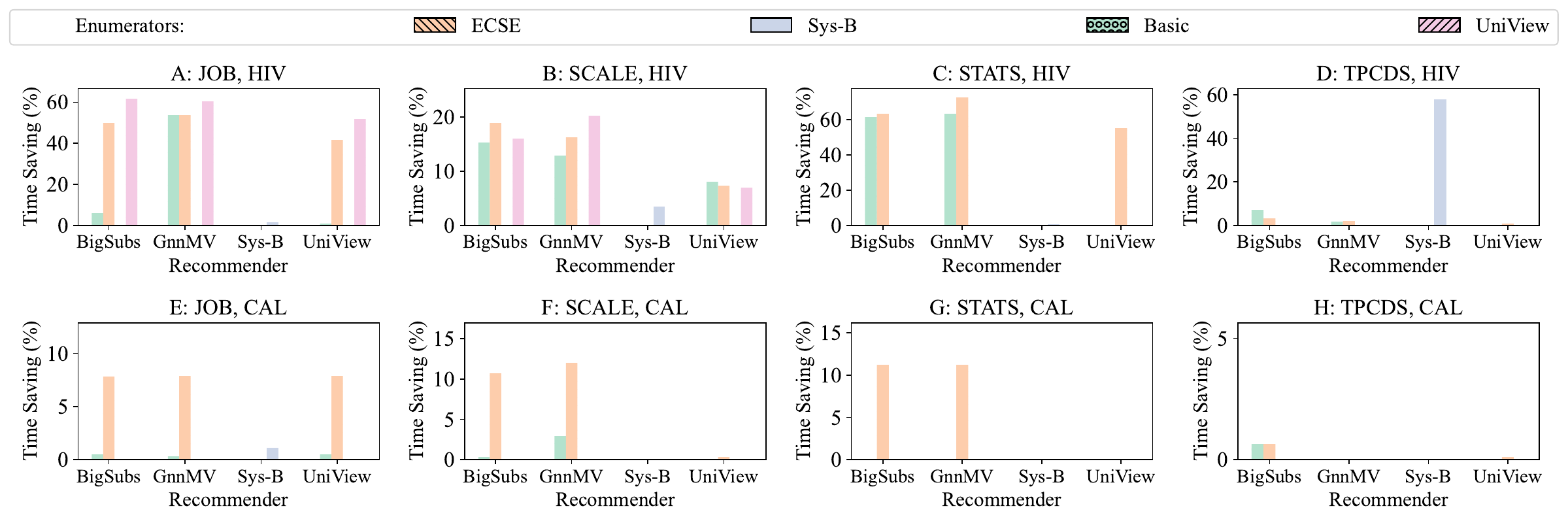}
    \vspace{-5mm}
    \caption{Workload time Saving (\%) of all (Enumerator,Selector,Rewriter) pipeline combinations on PostgreSQL.}
    \label{fig:pg_detailed_result}
\end{figure*}

\begin{figure}[]
    \centering
    \includegraphics[width=0.9\linewidth]{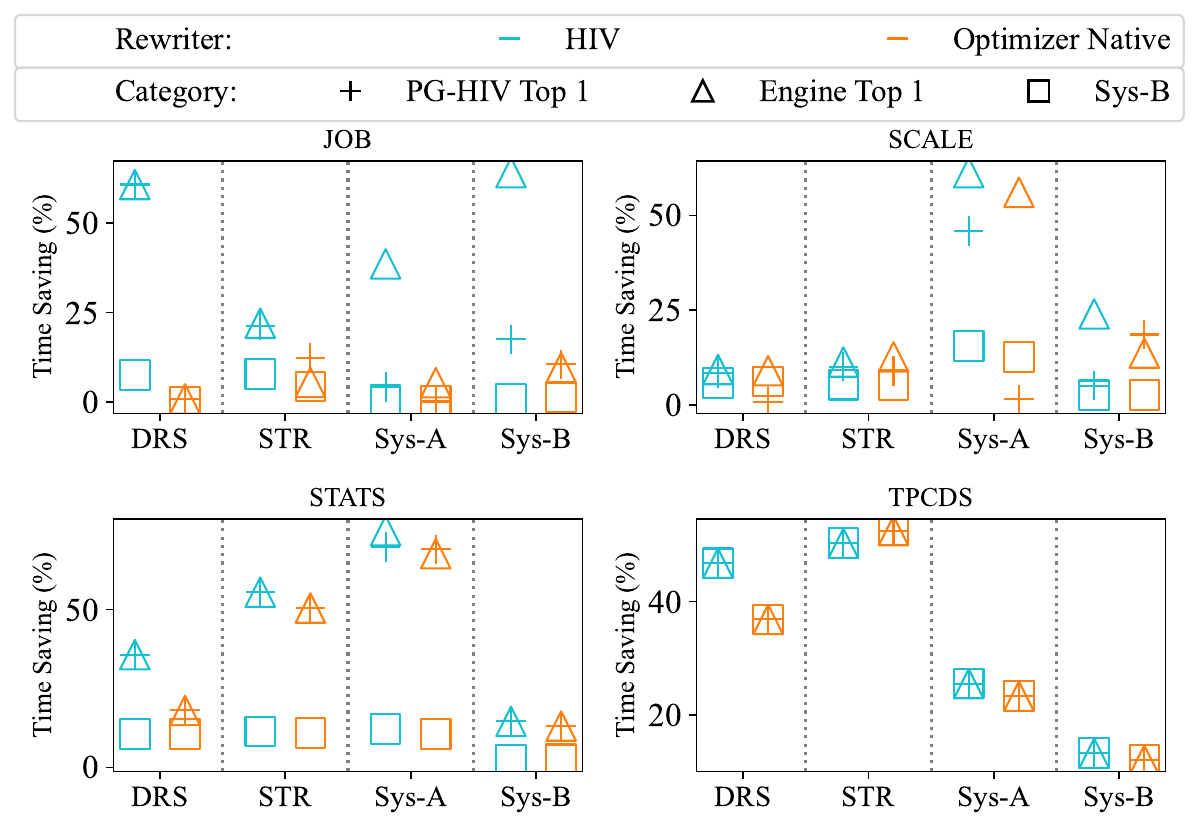}
    \vspace{-3mm}

    \caption{Cross-engine time saving under two rewrite modes. For each engine, three MV categories are examined to compare the performance of \underline{Optimizer Native} (let engine's native optimizer to rewrite) against \underline{\hive} (directly executing portable rewritten SQL generated by \hive). For MV categories, \underline{$+$}: ranking top 1 under PG-\hive{}, \underline{$\triangle$}: ranking top 1 under this engine, \underline{$\square$}: \celerdata{}'s auto MV selection. Budget=1GB.}
    \label{fig:exp-cross-engine}
\end{figure}

Figure~\ref{fig:pg_detailed_result} reports workload time saving for each pipeline on PostgreSQL with a fixed storage budget of 1 GB.

\noindent\textbf{PostgreSQL end-to-end: no consistent winner and strong interaction effects.} Across workloads, time savings vary substantially, and no single pipeline consistently dominates. For example, \ecse--\gnnmv--\hive{} achieves the largest savings on STATS, whereas \uniview--\bigsubs--\hive{} performs best on JOB. Even within the same workload, changing only one stage can lead to large differences: On JOB, replacing \ecse{} with \hawc{} in \ecse--\bigsubs--\hive{} reduces time saving by 43.85\,pp, whereas switching the selector from \bigsubs{} to \gnnmv{} under \hawc{} increases savings by 47.53\,pp, making it competitive with the best pipelines --- so stage rankings must be read \emph{jointly} with their upstream/downstream partners, not in isolation. We give the structural reason for this specific interaction (\bigsubs{}'s creation-cost-dominated utility under-ranking the high-coverage views \ecse{} contributes; \gnnmv{}'s learned cost recovering them under \hawc) in \textsection{\ref{sec:exp-enumerator}} and further by a detailed case study in \textsection\ref{sec:case-bigsubs-ranking}, and the corresponding workload-feature$\to$method-choice rule in the practical-guidance summary in \textsection\ref{sec:conclusion}. These results indicate strong interaction effects among candidate generation, selection under budget, and rewriting, supporting \textbf{RQ1} and motivating the stage-wise analyses in Sections~\ref{sec:exp-enumerator}--\ref{sec:exp-rewriter}.

\noindent \textbf{Cross-engine end-to-end: performance is engine and workload dependent.} We next evaluate cross-engine behavior using the best-performing modular pipelines identified on PostgreSQL under the cross-engine protocol. Across workloads, the best top-1 pipeline achieves 58--75\% time saving (JOB: 64\%, SCALE: 61\%, STATS: 75\%, TPC-DS: 58\%), well above the corresponding per-engine median; the full per-engine top-1 breakdown is reported in appendix. However, these top pipelines do not transfer uniformly across engines: the PostgreSQL top pipeline on JOB remains top-1 on Doris but yields smaller gains on StarRocks and limited gains on \redshift, while the PostgreSQL top pipeline on SCALE improves on \redshift~ but degrades on other engines. Overall, cross-engine MV benefits depend jointly on the selected view set, the rewriter's ability to exploit MVs, and engine-specific execution behavior.

When comparing portable \hive{} rewrites against each engine's native (optimizer-level) MV rewriting, we find that the best modular pipelines identified on PostgreSQL often yield larger time savings than the engine's native rewriter. In several cases the gap is substantial (e.g., JOB on Doris and \celerdata{}, and SCALE on \redshift{}; Figure~\ref{fig:exp-cross-engine}). On complex workloads such as JOB and STATS, the best modular pipelines achieve markedly higher time savings than the evaluated commercial systems in our setting (e.g., up to $\sim$60--70\% vs.\ $\sim$15--20\%; Figure~\ref{fig:exp-cross-engine}). In contrast, on TPC-DS the gap narrows and, in some cases, commercial systems perform the best. Overall, these results indicate that MV-based query rewriting effectiveness is highly workload-dependent and engines exhibit different trade-offs across workloads (\textbf{RQ2}). We investigate the sources of these differences in the subsequent stage-wise analyses, including cases where candidate generation, view selection under budget, or rewrite applicability becomes the limiting factor.

\begin{figure}[b]
    \centering
    \includegraphics[width=\linewidth]{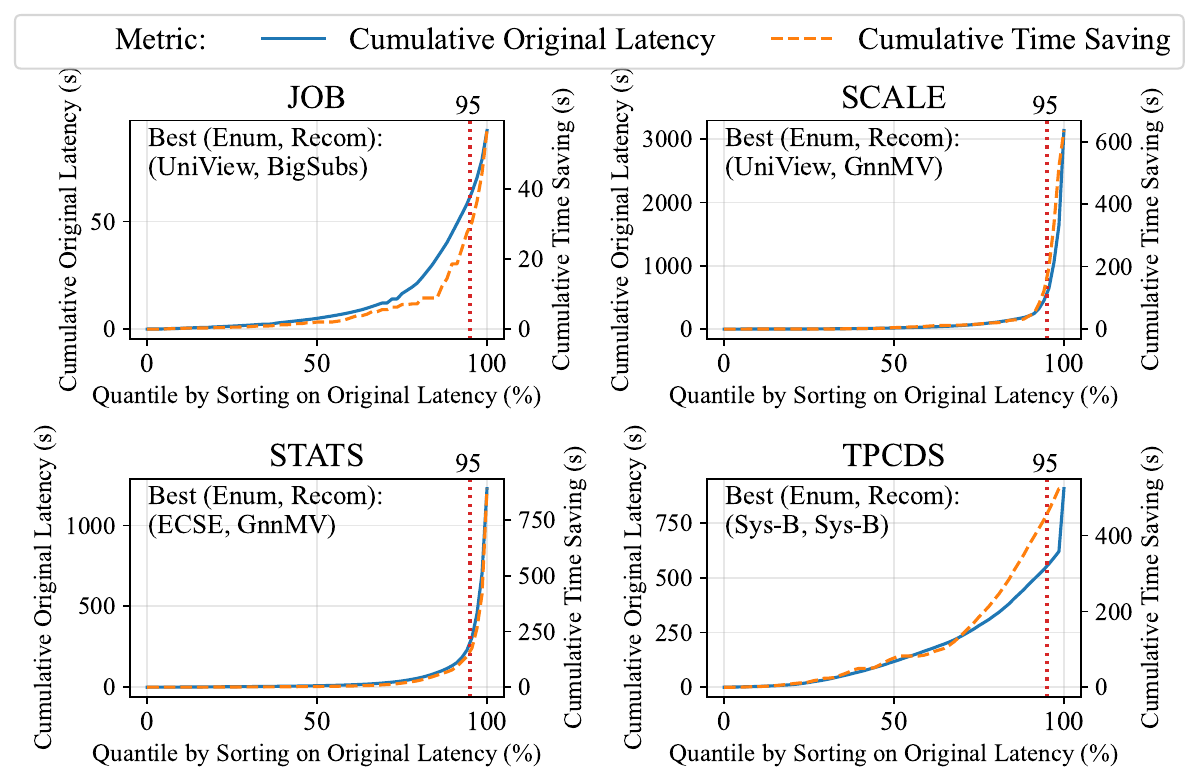}
    \vspace{-8mm}
    \caption{Cumulative Original Latency and Time Saving Distribution by Quantile. Queries are sorted by original latency in ascending order to determine the quantile.}
    \label{fig:cumulative_ts_by_quantile}
\end{figure}

\noindent \textbf{Overall performance is determined by long tail minorities.} We sorted queries by original latency and plotted the cumulative original latency and time saving achieved by the best combination (Figure \ref{fig:cumulative_ts_by_quantile}). Each vertical dashed line separates the top 5\% slowest queries from the rest. The (original latency, time saving) shared contributions by the slowest queries are: ($36.93\%$, $51.42\%$) for \job, ($81.15\%$, $71.73\%$) for \scale, ($77.69\%$, $82.62\%$) for \stats~ and ($39.13\%$, $12.97\%$) for \tpcds. For those workloads where the best combinations are all non-commercial combinations (\job, \scale, \stats), the slowest queries not only contribute the most time saving, but their magnitude is also comparable to the original latency contribution. Only in \tpcds~ where the \celerdata~ pipeline is chosen, is the share of time saving by the slowest queries much smaller than original latency contribution. That indicates although \celerdata~ achieves the best performance in \tpcds, it does not optimize the slowest queries very well in proportion to the rest of the queries.

\section{Stage-wise Analyses}
\label{sec:stage_wise_analysis}
The end-to-end results exhibit strong interaction effects. We therefore perform controlled ablations: when analyzing one stage, we fix the other two stages to isolate its contribution.

\subsection{ Analysis of Enumerators}
\label{sec:exp-enumerator}

\noindent\textbf{Enumerators differ in candidate count, coverage, and structure.} The best enumerator varies by workload (\ecse/\uniview{} on \job/\scale, \ecse{} on \stats, \celerdata{}\footnote{\celerdata{} exposes views only after its internal selection.} on \tpcds), indicating that enumerators differ not just in how many candidates they generate but in the \emph{types} of views they expose. We analyze candidate count, coverage, and structure to explain these differences.

\begin{table}[t]
\centering
\caption{ Candidate-space statistics. Coverage is the fraction of workload queries covered by each candidate view. JO: number of unique \underline{j}oin \underline{o}rders; J: mean number of \underline{j}oins; W: mean number of unique \texttt{\underline{W}HERE} operators.}
\label{tab:enum_candidate_stats}
\vspace{-2mm}
\scriptsize
\setlength{\tabcolsep}{2.2pt}
\begin{tabular}{llcrrrrrrrr}
\toprule
\multirow{2}{*}{\textbf{Workload}} 
& \multirow{2}{*}{\textbf{Enumerator}} 
& \multirow{2}{*}{\textbf{\# Candidates}} 
& \multicolumn{4}{c}{\textbf{Coverage (\%)}} 
& \multicolumn{3}{c}{\textbf{Complexity}} \\
\cmidrule(lr){4-7}\cmidrule(lr){8-10}
& & 
& \textbf{Avg} 
& \textbf{q50} 
& \textbf{q95} 
& \textbf{Max} 
& \textbf{JO} 
& \textbf{J} 
& \textbf{W} \\
\midrule

\multirow{4}{*}{\job{}} 
& \celerdata{} & 11  & 3.31  & 4.55  & 4.55  & 4.55  & 5   & 5.27 & 0.64 \\
& \ecse{}      & 373 & 14.84 & 9.09  & 38.64 & 70.45 & 243 & 2.34 & 0.06 \\
& \hawc{}      & 146 & 19.52 & 18.18 & 56.82 & 61.36 & 78  & 2.47 & 0.37 \\
& \uniview{}   & 95  & 3.78  & 2.27  & 13.64 & 15.91 & 43  & 2.13 & 7.98 \\
\midrule

\multirow{4}{*}{\scale{}} 
& \celerdata{} & 70  & 1.44 & 1.00 & 4.28  & 6.00  & 18 & 1.30 & 0.00 \\
& \ecse{}      & 566 & 3.89 & 2.00 & 13.00 & 39.50 & 30 & 1.42 & 5.27 \\
& \hawc{}      & 569 & 3.38 & 1.50 & 11.50 & 39.50 & 56 & 1.41 & 5.41 \\
& \uniview{}   & 419 & 1.29 & 0.50 & 4.50  & 36.00 & 34 & 1.58 & 7.69 \\
\midrule

\multirow{3}{*}{\stats{}} 
& \celerdata{} & 24  & 0.84 & 0.69 & 1.56 & 1.73  & 19 & 2.83 & 0.00 \\
& \ecse{}      & 394 & 2.38 & 0.87 & 9.34 & 47.23 & 45 & 2.44 & 5.36 \\
& \hawc{}      & 392 & 1.34 & 0.69 & 3.88 & 13.49 & 32 & 2.50 & 5.80 \\
\midrule

\multirow{3}{*}{\tpcds{}} 
& \celerdata{} & 25  & 3.49 & 3.48 & 6.52 & 6.68  & 18 & 3.32 & 0.00 \\
& \ecse{}      & 410 & 1.28 & 0.27 & 5.61 & 15.78 & 54 & 2.30 & 5.99 \\
& \hawc{}      & 305 & 0.77 & 0.27 & 3.96 & 7.22  & 19 & 2.44 & 7.23 \\
\bottomrule
\end{tabular}

\footnotesize
\end{table}

\noindent \textbf{Candidate-space and reuse potential.} Table~\ref{tab:enum_candidate_stats} reports candidate counts and the per-view covered-query ratio. \ecse{} and \hawc{} enumerate far more candidates than \uniview{}, and \celerdata{} the fewest; the \ecse--\hawc{} gap widens on the most complex workload (\job{}).

Across workloads, candidate count does not directly imply higher reuse: on \job, \ecse{} enumerates 373 candidates (vs.\ 146 for \hawc{}) but has lower median and q95 coverage (q50: 9.09\% vs.\ 18.18\%; q95: 38.64\% vs.\ 56.82\%). In contrast, on \stats{} and \tpcds{}, \ecse{}  produces substantially higher tail coverage than \hawc~ (e.g., max 47.23\% vs.\ 13.49\% on \stats; max 15.78\% vs.\ 7.22\% on \tpcds), indicating enumeration strategy can shift reuse from the ``typical'' candidate to a small set of ``high-leverage'' candidates. A consistent pattern across datasets is that coverage is skewed: most candidates cover few queries, while a small fraction achieve very high coverage (e.g., the max coverage reaches 70.45\% on \job{}). Whether downstream selection actually exploits this heavy-tailed structure depends on the selector's utility model: we show in \textsection\ref{sec:case-bigsubs-ranking} that \bigsubs{}'s creation-cost-dominated utility under-ranks exactly the high-coverage views in this tail on \job{}, whereas \gnnmv{}'s learned cost prioritizes them --- so the same enumerated candidate pool yields different downstream behavior depending on selector choice (the 47.53\,pp difference under \hawc{} reported in \textsection\ref{sec:exp-endtoend}).

\noindent \textbf{Mechanism: structural bias, not candidate count.}
\ecse{} expands the space mainly through \emph{join-structure variants} --- on \job{} it produces 243 distinct join orders vs.\ \hawc{}'s 78 and \uniview{}'s 43 at comparable join counts (Table \ref{tab:enum_complexity} in appendix) --- which is why its high-coverage candidates concentrate in the tail. Enumerators also differ in predicate richness (\uniview{} $\sim$8 \texttt{WHERE} operators per view vs.\ \celerdata{}'s $\sim$0), so the same selector and rewriter can yield very different end-to-end savings across enumerators.

\noindent \textbf{Candidate structural characteristics.}
Table~\ref{tab:enum_complexity} (in appendix) summarizes structural properties of the enumerated candidates. 
On \job{}, \ecse{} generates far more distinct join orders than other enumerators (243 vs.\ 78 for \hawc{} and 43 for \uniview{}), while the average join count is comparable (2.34 vs.\ 2.47/2.13). Combined with the coverage per view, this indicates that \ecse{} expands the candidate space primarily through \emph{structural variants} rather than enumerating larger join subexpressions, and helps explain why \ecse{} contains higher-\emph{tail}-coverage candidates on \job{}.

\begin{figure}[t]
    \centering
    \includegraphics[width=\linewidth]{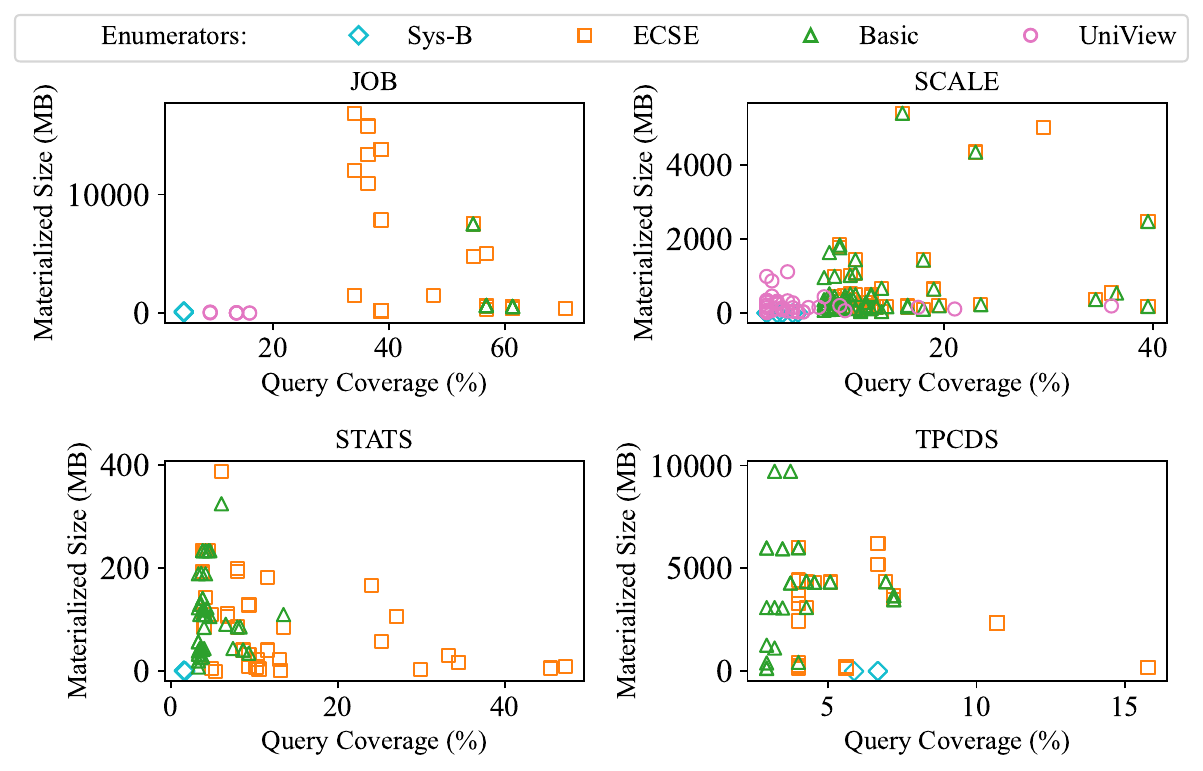}
    \vspace{-8mm}
    \caption{Query coverage vs. materialized size for the top 10\% highest query coverage views from each enumerator. Size measured on PostgreSQL.}
    \label{fig:query_coverage_vs_size}
\end{figure}

\noindent\textbf{Coverage vs. size: high reuse may not imply large views.}
A common intuition is that high-coverage views must be more general (fewer predicates, more projected columns) and therefore larger in size. Figure~\ref{fig:query_coverage_vs_size} tests this by plotting the materialized size distribution of the top 10\% candidates ranked by coverage for each enumerator (Table~\ref{tab:enum_candidate_stats}).
We find that the size--coverage trade-off is not universal: on \job{}, \stats{}, and \tpcds{}, \ecse{} produces many high-coverage candidates that are also relatively small, whereas other enumerators more closely follow the expected trade-off. On \scale{}, where query structure is simpler, \ecse{} behaves similarly to \hawc{}. Overall, these results suggest that for complex workloads, \ecse{} can expose more space-effective candidates (high reuse with modest materialization size), which helps downstream selection under budget.

\RCOMMENT{
\noindent\textbf{Takeaways.} \textbf{(RQ1/RQ3)} Enumeration sets the achievable upper bound on end-to-end savings --- a \emph{cap}, not a \emph{cause}. Across the end-to-end pipeline combinations evaluated in \textsection\ref{sec:exp-endtoend} (Figure~\ref{fig:pg_detailed_result}), no downstream selector or rewriter recovers the savings lost when the enumerator omits a high-coverage candidate: e.g., \hawc's top-1 candidate on \job{} covers 18.18\% of queries vs.\ \ecse{}'s 70.45\%, and no selector--rewriter pair on the \hawc{} candidate set matches the \ecse-best pipeline's end-to-end time saving on \job.

\noindent \textbf{(RQ3)} Candidate count is not a proxy for usefulness: larger candidate spaces do not necessarily yield higher reuse or savings; reuse (coverage) is typically \emph{heavy-tailed}, driven by a small set of high-impact views.

\noindent \textbf{(RQ3)} Structural bias matters: enumerators differ in join-order exploration and predicate specialization, which shifts where reuse appears and affects which queries downstream stages can accelerate.
}

\subsection{Analysis of Selectors}
\label{section:exp-selector}

\noindent \textbf{Selectors differ more in robustness than in average performance.}
In the end-to-end results (Figure~\ref{fig:pg_detailed_result}), no selector wins universally: the ranking of \bigsubs{}, \gnnmv{}, and \uniview{} changes across workloads and (enumerator, rewriter) settings. We therefore study selector robustness through a storage-budget sweep over the three budget-aware selectors, excluding \celerdata{} because its integrated recommender does not expose an explicit storage-budget interface.

\begin{figure}
    \centering
\includegraphics[width=0.9\linewidth]{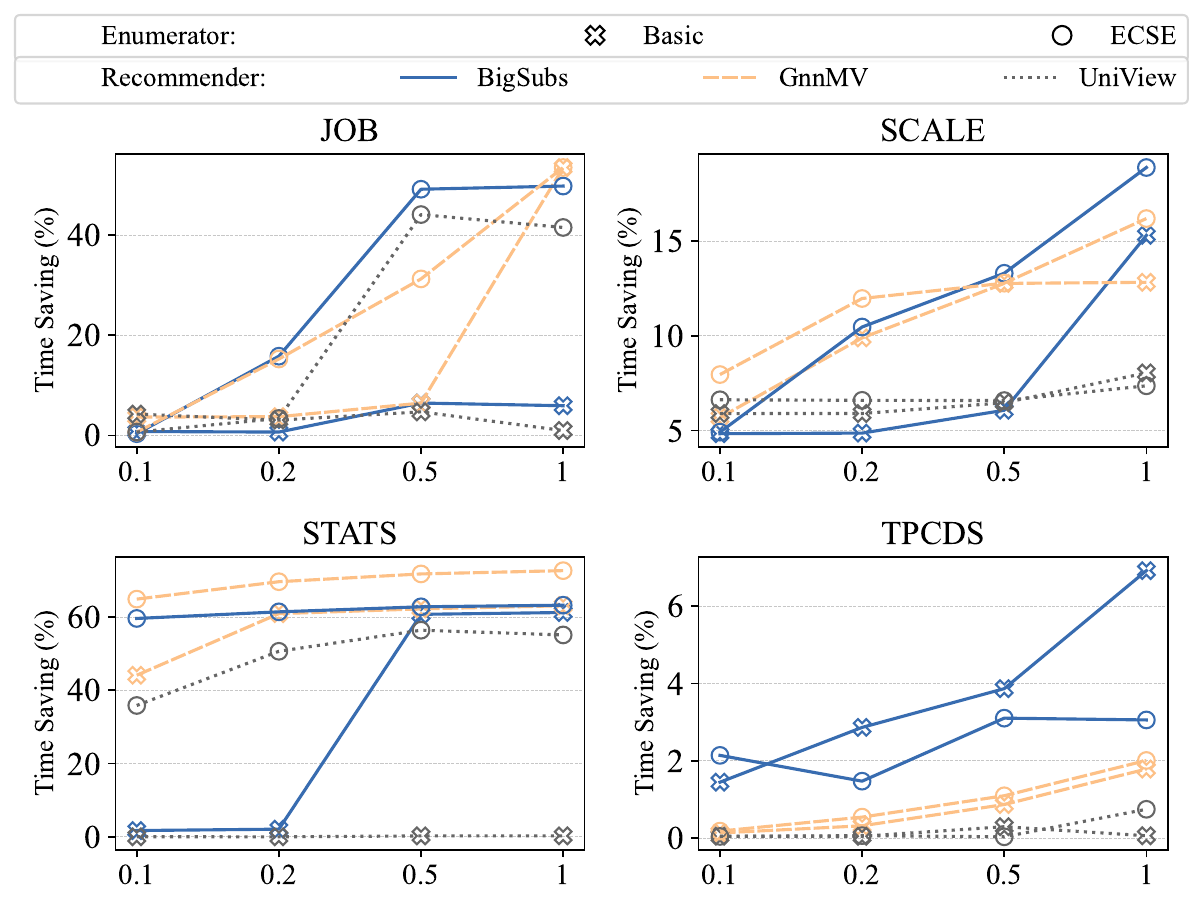}
\vspace{-3mm}
    \caption{Time Saving of Each Enumerator-Recommender Combination in Different Storage Budgets}
    \label{fig:ts_by_budget}
\end{figure}

\noindent\textbf{Budget sensitivity and interaction with enumerators.}
Figure~\ref{fig:ts_by_budget} varies the storage budget (0.1/0.2/0.5/1\,GB) under two enumerators (\ecse{} and \hawc{}), while fixing rewriter to \hive{} and engine to \pg{}. 
Overall, selector behavior is highly workload-dependent. On \job{}, \scale{}, and \stats{}, \gnnmv{} is usually best or near-best across budgets, while \bigsubs{} is competitive but occasionally yields limited savings; \uniview{} generally underperforms. On \tpcds{}, \bigsubs{} performs best, but the absolute savings are smaller, indicating fewer exploitable MV opportunities under this budget range.

Two patterns explain why selector choice matters. First, savings often show threshold effects: moving from 0.2 to 0.5\,GB or from 0.5 to 1\,GB can produce abrupt gains, suggesting that a few high-impact views sit near budget boundaries. Second, selectors interact strongly with enumerators. For example, on \job{}, \hawc{}-based pipelines perform poorly with \bigsubs{} and \uniview{}, but become competitive when paired with \gnnmv{} at 1\,GB. Thus, a selector can sometimes compensate for a weaker candidate set by prioritizing the remaining high-impact views.

\noindent \textbf{Diagnosing cases with large selector gaps.}
To understand when selector matters most, we drill down into the outlier where \gnnmv{} substantially outperforms \bigsubs{} under a tight budget. Since both methods saturate the storage budget and are cost-aware, the gap cannot be explained by budget under-utilization; instead, it reflects differences in \emph{which} views are prioritized.
We find evidence that \bigsubs{} can under-rank certain high-impact views due to its simplified utility model, leading to suboptimal selections when benefits depend on factors that are not well captured by the estimator (\textsection \ref{sec:case-bigsubs-ranking}).

\noindent\textbf{Latency overhead.}
Selection overhead is consistent across workloads: \bigsubs{} is cheapest (no training; sub-second ILP), \gnnmv{} adds 288--1{,}551\,s of training, and \uniview{} is the most expensive; the integrated \celerdata{} scales more gently with workload size ($2.27\times$ from JOB to STATS vs.\ $\sim$5.3--5.5$\times$ for the learned selectors). Full numbers are in appendix.

\RCOMMENT{
\noindent\textbf{Takeaways.} \textbf{(RQ1)} Learned selectors yield more stable performance across budgets on complex workloads, with moderate training/inference overhead, compared to ILP-based \bigsubs{}.
    
\noindent \textbf{(RQ3)} Budget sensitivity is significant: selector rankings can flip between 0.1–0.2 GB and 0.5–1 GB, and savings often exhibit threshold effects when a small set of high-impact views becomes feasible.

\noindent \textbf{(RQ3)} Under tight budgets, performance is driven primarily by \emph{view ranking} rather than budget utilization; a cost-based utility model can under-rank a few high-impact views.
}
\subsection{Analysis of Rewriter}
\label{sec:exp-rewriter}

\noindent \textbf{Rewriter capability often caps end-to-end time savings.}
Regarding \textbf{RQ3}, we find that the rewriter frequently determines how much benefit upstream enumeration and selection can translate into actual workload savings. With the enumerator and selector fixed, replacing the rewriter can change the time saving substantially. For example, on \job{} with the same selected view set (\hawc{} + \gnnmv{}), switching from \hive{} to \redshift{}'s native rewriter reduces workload time saving from $\sim$40\% to $\sim$5\% (Figure~\ref{fig:exp-cross-engine}). This suggests that, in many scenarios, the limiting factor is not only cost estimation (since \hive{} does not require physical materialization to produce a rewrite) or view ranking, but whether the rewriter can \emph{recognize} applicable views and \emph{apply} effective substitutions. 
We further analyze rewrite success, view usage, and representative failure modes below.

{
\begin{table}[htbp]\centering
\scriptsize
\caption{Rewrite success rate (\%) and count of successful rewrites for each rewriter. Green and yellow shading mark the highest success rate among \protect\fancycellGreenNormal{non-commercial} and \protect\fancycellYellowNormal{commercial} rewriters, respectively. Bold entries denote the overall highest success rate across all rewriters.}
\label{tab:exp-rewrite-success}
\vspace{-2mm}
\begin{tabular}{C{0.45cm}C{0.5cm}C{0.5cm}C{0.7cm}C{0.7cm}C{0.65cm}C{0.6cm}C{0.6cm}C{0.6cm}}\toprule
\textbf{WL} &\textbf{Enum.} &\textbf{CAL} &\textbf{Sys-A} &\textbf{Sys-B} &\textbf{Sys-C} &\textbf{DRS} &\textbf{HIV} &\textbf{STR} \\\midrule
\multirow{4}{*}{JOB} &UniView &\makecell{4\\ (2\%)} &\makecell{31\\ (15\%)} &\makecell{\fancycellYellow{146}\\ \fancycellYellow{(71\%)}} 
& \makecell{2\\ (1\%)}
&\makecell{7\\ (3\%)} &\makecell{\fancycellGreen{\textbf{152}}\\ \fancycellGreen{\textbf{(73\%)}}} &\makecell{146\\ (71\%)} \\
&ECSE &\makecell{4\\ (7\%)} &\makecell{\fancycellYellow{\textbf{49}}\\ \fancycellYellow{\textbf{(80\%)}}} &\makecell{15\\ (25\%)} 
& \makecell{1\\ (2\%)}
& \makecell{0\\ (0\%)}&\makecell{\fancycellGreen{39}\\ \fancycellGreen{(64\%)}} &\makecell{26\\ (43\%)} \\
&Sys-B &\makecell{5\\ (26\%)} &\makecell{\fancycellYellow{16}\\ \fancycellYellow{(84\%)}} &\makecell{\fancycellYellow{16}\\ \fancycellYellow{(84\%)}} 
& \makecell{4\\ (21\%)}
& \makecell{0\\ (0\%)}&\makecell{\fancycellGreen{\textbf{19}}\\ \fancycellGreen{\textbf{(100\%)}}} &\makecell{16\\ (84\%)} \\
&Basic &\makecell{6\\ (10\%)} &\makecell{\fancycellYellow{\textbf{50}}\\ \fancycellYellow{\textbf{(79\%)}}} &\makecell{11\\ (17\%)} 
& \makecell{5\\ (8\%)}
&\makecell{2\\ (3\%)} &\makecell{\fancycellGreen{\textbf{50}}\\ \fancycellGreen{\textbf{(79\%)}}} &\makecell{14\\ (22\%)} \\\hline
\multirow{4}{*}{SCALE} &UniView & \makecell{0\\ (0\%)}&\makecell{15\\ (3\%)} &\makecell{\fancycellYellow{\textbf{404}}\\ \fancycellYellow{\textbf{(87\%)}}} 
& \makecell{45\\ (10\%)}
&\makecell{23\\ (5\%)} &\makecell{186\\ (40\%)} &\makecell{\fancycellGreen{\textbf{404}}\\ \fancycellGreen{\textbf{(87\%)}}} \\
&ECSE &\makecell{51\\ (4\%)} &\makecell{47\\ (4\%)} &\makecell{\fancycellYellow{1231}\\ \fancycellYellow{(94\%)}} 
& \makecell{49\\ (4\%)}
&\makecell{51\\ (4\%)} &\makecell{78\\ (6\%)} &\makecell{\fancycellGreen{\textbf{1243}}\\ \fancycellGreen{\textbf{(95\%)}}} \\
&Sys-B & \makecell{0\\ (0\%)}&\makecell{\fancycellYellow{\textbf{281}}\\ \fancycellYellow{\textbf{(100\%)}}} &\makecell{\fancycellYellow{\textbf{281}}\\ \fancycellYellow{\textbf{(100\%)}}} 
& \makecell{0\\ (0\%)}
&\makecell{\fancycellGreen{\textbf{281}}\\ \fancycellGreen{\textbf{(100\%)}}} &\makecell{\fancycellGreen{\textbf{281}}\\ \fancycellGreen{\textbf{(100\%)}}} &\makecell{\fancycellGreen{\textbf{281}}\\ \fancycellGreen{\textbf{(100\%)}}} \\
&Basic &\makecell{76\\ (5\%)} &\makecell{116\\ (8\%)} &\makecell{\fancycellYellow{1413}\\ \fancycellYellow{(93\%)}} 
& \makecell{63\\ (4\%)}
&\makecell{104\\ (7\%)} &\makecell{152\\ (10\%)} &\makecell{\fancycellGreen{\textbf{1420}}\\ \fancycellGreen{\textbf{(93\%)}}} \\\hline
\multirow{3}{*}{STATS} &ECSE &\makecell{110\\ (3\%)} &\makecell{\fancycellYellow{2552}\\ \fancycellYellow{(78\%)}} &\makecell{2237\\ (69\%)} 
& \makecell{260\\ (8\%)}
&\makecell{747\\ (23\%)} &\makecell{\fancycellGreen{\textbf{3200}}\\ \fancycellGreen{\textbf{(98\%)}}} &\makecell{2214\\ (68\%)} \\
&Sys-B & \makecell{0\\ (0\%)} &\makecell{\fancycellYellow{\textbf{230}}\\ \fancycellYellow{\textbf{(100\%)}}} &\makecell{\fancycellYellow{\textbf{230}}\\ \fancycellYellow{\textbf{(100\%)}}} 
& \makecell{0 \\(0\%)}
&\makecell{224\\ (97\%)} &\makecell{\fancycellGreen{\textbf{230}}\\ \fancycellGreen{\textbf{(100\%)}}} &\makecell{\fancycellGreen{\textbf{230}}\\ \fancycellGreen{\textbf{(100\%)}}} \\
&Basic &\makecell{54\\ (5\%)} &\makecell{661\\ (66\%)} &\makecell{\fancycellYellow{788}\\ \fancycellYellow{(79\%)}} 
& \makecell{145\\ (15\%)}
&\makecell{389\\ (39\%)} &\makecell{\fancycellGreen{\textbf{887}}\\ \fancycellGreen{\textbf{(89\%)}}} &\makecell{775\\ (78\%)} \\\hline
\multirow{3}{*}{TPCDS} &ECSE &\makecell{14\\ (12\%)} &\makecell{\fancycellYellow{97}\\ \fancycellYellow{(83\%)}} &\makecell{61\\ (52\%)} 
& \makecell{49\\ (42\%)}
&\makecell{\fancycellGreen{\textbf{114}}\\ \fancycellGreen{\textbf{(97\%)}}} &\makecell{103\\ (88\%)} &\makecell{61\\ (52\%)} \\
&Sys-B & \makecell{0\\ (0\%)}&\makecell{\fancycellYellow{\textbf{469}}\\ \fancycellYellow{\textbf{(100\%)}}} &\makecell{\fancycellYellow{\textbf{469}}\\ \fancycellYellow{\textbf{(100\%)}}} 
& \makecell{0 \\ (0\%)}
&\makecell{303\\ (65\%)} &\makecell{\fancycellGreen{\textbf{469}}\\ \fancycellGreen{\textbf{(100\%)}}} &\makecell{\fancycellGreen{\textbf{469}}\\ \fancycellGreen{\textbf{(100\%)}}} \\
&Basic &\makecell{14\\ (16\%)} &\makecell{\fancycellYellow{71}\\ \fancycellYellow{(83\%)}} &\makecell{58\\ (67\%)} 
& \makecell{67\\ (78\%)}
&\makecell{59\\ (69\%)} &\makecell{\fancycellGreen{\textbf{86}}\\ \fancycellGreen{\textbf{(100\%)}}} &\makecell{58\\ (67\%)} \\
\bottomrule
\end{tabular}
\end{table}
}

\noindent \textbf{Rewrite success rate varies across engines and workloads.}
Table~\ref{tab:exp-rewrite-success} reports rewrite success rates under each rewriter, grouped by workload and enumerator.\footnote{We measure success rate as the fraction of query--view pairs for which the rewriter produces a query/plan that references an MV, under a fixed pool of MVs. 
}
Across \job{}, \stats{}, and \tpcds{}, \hive{} is the most consistently effective rewriter, achieving the highest success rates in 9 out of 14 workload--enumerator settings; \starrocks{} and \doris{} lead in fewer cases. Among commercial systems, \redshift{} more frequently exhibits higher rewrite-success rates than \celerdata{} in these workloads. \bigquery{} exhibits the most variable behavior across workloads: only 1--2\% rewrite success on \job{} (limited support for the workload's complex multi-join patterns), 4--15\% on \scale{} and \stats{}, but 42--78\% on \tpcds{}, where its MV-rewriter handles aggregate-heavy templates well. However, there is no universal winner: in some settings (e.g., parts of \stats{} and \tpcds{}), multiple systems reach near-100\% success rates, indicating that these workloads contain patterns well covered by several optimizers. In contrast, on \scale{} the ranking changes: \starrocks{} becomes the top-performing open-source rewriter and \celerdata{} becomes the top-performing commercial rewriter, with comparable success rates. 

Overall, these results reinforce that rewriter behavior is strongly workload- and engine-dependent, and can be a dominant source of end-to-end variability even when the view set is fixed. We conduct a case study in Section~\ref{sec:case-rewriter} (appendix) to investigate why \hive{} achieves higher success rates in most settings, and compare \hive{} with \doris{} in depth on \stats{}. Additionally, via a case study in Section \ref{sec:risky_rewrite} in appendix, we show that MVs can also be "forced" to be used to rewrite queries under certain engines, which in turn does not contribute to performance, and sometimes we can easily detect those risky rewrites.

\begin{figure*}[htp]\captionsetup[subfigure]{font=footnotesize}
\centering
\vspace{-3mm}
\begin{subfigure}{0.49\linewidth}
\centering
\caption{\job~}
    \includegraphics[width=\linewidth]{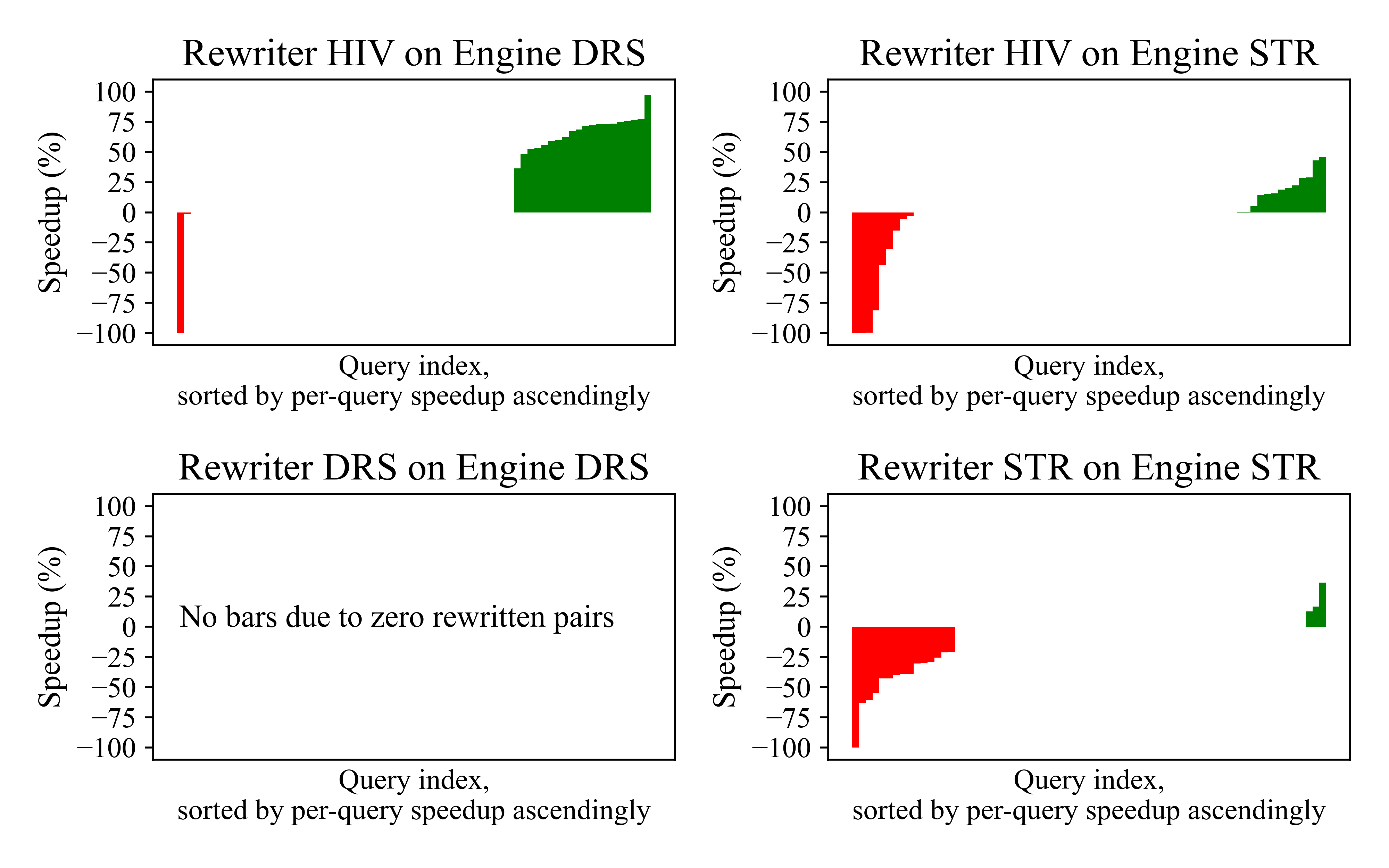}
    \label{fig:ts_main_result_tpcds}
\end{subfigure}
\hfill
\begin{subfigure}{0.49\linewidth}
\centering
\caption{\scale~}
    \includegraphics[width=\linewidth]{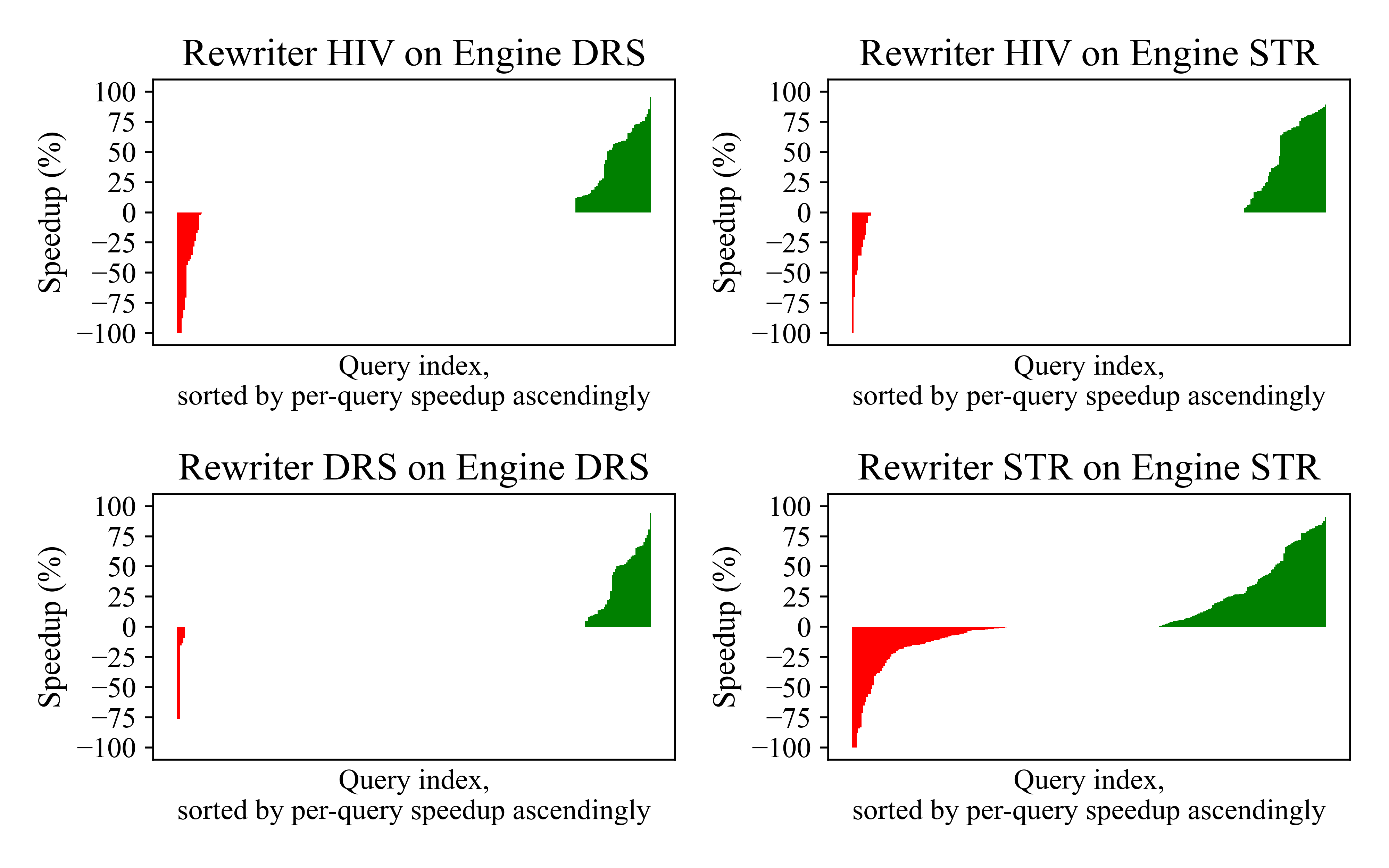}
    \label{fig:ts_main_result_scale}
\end{subfigure}%

\caption{Per-query runtime impact of each rewriter on \job{} and \scale{}. Each bar is one successfully rewritten query; bar height is the relative speedup. Positive (accelerated) and negative (slowdown) speedup are marked as green and red respectively. Queries on the X-axis are sorted left-to-right by its speedup magnitude; the ordering is independent per subplot.} \label{fig:exp-rewriter-ts-distribution}

\label{fig:exp-rewriter-ts-distribution}
\label{fig:pos_neg_speedup_by_rewriter}
\end{figure*}
\noindent\textbf{Runtime impact varies substantially, revealing different robustness profiles.}
Figure~\ref{fig:exp-rewriter-ts-distribution} presents the rewrite impact on each query on \job{} and \scale{}. Each vertical bar corresponds to a successfully rewritten query, and its height is the relative speedup (positive) or slowdown (negative) compared to executing the original query on the same engine. The distributions expose substantial heterogeneity within a single workload: (i) many successful rewrites yield large gains, but a non-trivial fraction produce regressions, showing that rewrite ``success'' does not imply performance improvement; (ii)  many rewrites deliver limited or even zero benefit, e.g., \hive{} and \starrocks{} on JOB, \hive{} and \doris{} on \scale{}; (iii) rewriters differ not only in the fraction of positive pairs, but also in tail behavior, e.g., some exhibit both heavy negative and positive tails (severe regressions as well as large improvements, e.g., \hive{} and \starrocks{}  for \scale{}), whereas others exhibit non-symmetric tails (e.g.,  \hive{} on Doris engine for \job{}, \starrocks{} for \job{}).

\RCOMMENT{\noindent\textbf{Takeaways.} \textbf{(RQ2)} Rewriter behavior is \emph{engine- and workload-dependent}, with no universal winner. Under a fixed view set (Table~\ref{tab:exp-rewrite-success}), \hive{} achieves the highest rewrite-success rate in 9 of 14 workload--enumerator settings across \job/\stats/\tpcds{}, yet \starrocks{} leads on \scale{} and several systems tie near 100\% elsewhere, supporting auto-MV rewrite does not imply consistent rewrite usage.

\noindent \textbf{(RQ1/RQ2)} Rewrite \emph{success} is decoupled from \emph{benefit}, and robustness lives in the \emph{tails}. Under data-distribution drift, \hive{}'s rewrite-success rate stays nearly flat ($89\%\to91\%$) while \doris{}'s collapses ($99\%\to28\%$) on the same \ecse-enumerated view set (Table~\ref{tab:skewed_distribution_rewritability} in appendix); and on \job{}, switching only the rewriter (\hive$\to$\redshift) under a fixed view set cuts time saving $\sim$40\%$\to\sim$5\% (Figure~\ref{fig:exp-cross-engine}). Rare but severe negative-speedup tails on specific (rewriter, engine) pairs can therefore dominate workload-level outcomes.

\noindent \textbf{(RQ2)} Under the cross-engine protocol, \hive{}-rewritten SQL provides a controlled baseline against native optimizer-level rewriting, and its advantage is workload-dependent: \hive{} dominates plan-transparent rewriters on \stats-\ecse{} (98\% vs.\ 23--78\%) but trails on \tpcds-\ecse{} (88\% vs.\ \doris{} 97\%) (Table~\ref{tab:exp-rewrite-success})---most useful where engine-native rewriting is weakest.
}
\subsection{Robustness Analysis}
\label{sec:robustness-summary}

The preceding analyses use a controlled steady-state setting: the MV-construction workload matches the evaluation workload, the database instance is fixed, and the same resource environment is used throughout. We add three robustness checks beyond this: workload drift, data-distribution drift, and memory pressure. Table~\ref{tab:robustness_summary} summarizes the main results. We elaborate the workload-drift analysis here because it directly tests the representativeness assumption of offline MV construction; full data-skew and hardware-pressure case studies are provided in appendix.

\begin{table}[htbp]
\centering
\caption{Summary of robustness experiments. Full data-skew and hardware-pressure results are in appendix.}
\label{tab:robustness_summary}
\vspace{-2mm}
\scriptsize
\setlength{\tabcolsep}{2.5pt}
\begin{tabular}{p{1.45cm}p{1.9cm}p{2.2cm}p{2.0cm}}
\toprule
\textbf{Stress} & \textbf{What changes} & \textbf{Key result} & \textbf{Takeaway} \\
\midrule
Workload drift (\textsection \ref{sec:robustness-drift})
& MV construction workload differs from evaluation workload on \imdb{}.
& \ecse{} is stable (shrink factors 1.38/0.73), while \uniview{} is sensitive (3.18/$+\infty$).
& Join-graph candidates transfer better than predicate-specialized candidates. \\
\midrule
Data-distribution drift (\textsection \ref{sec:data_skewness_short} and \textsection \ref{sec:robustness-data} in appendix)
& Same \tpcds{} templates; balanced \tpcds{} vs.\ skewed \dsb{} instance.
& On skew-sensitive \pg{} queries, \gnnmv{} drops 71.67\%$\to$49.71\%, while \bigsubs{} drops 61.04\%$\to$53.38\%.
& Skew affects both realized speedup and selector robustness. \\
\midrule
Hardware / memory pressure
& MV recommendation and evaluation run across abundant vs.\ constrained EC2 settings.
& Recommended MV sets often remain useful across environments, but native rewrite success can change for some engines.
& MV selection transfers reasonably across resources; rewriting remains engine-sensitive. \\
\bottomrule
\end{tabular}
\end{table}

\subsubsection{Workload Drift}
\label{sec:robustness-drift}

Workload drift tells whether the workload used for MV construction is representative of the workload used for evaluation. We study this using two workloads (\job{} and \scale{}) with the rewriter fixed to \hive{}, the execution engine to \pg{}, and the storage budget to 1\,GB. We report relative workload time saving and average results across selectors. The shrink factor is computed as the matched-setting saving divided by the mismatched-setting saving; larger values indicate stronger degradation under drift.

\begin{table}[!htp]
\centering
\small
\caption{Time saving (\%) under workload drift. Shrink factor is computed as \colorbox[HTML]{D9D2E9}{matched} / \colorbox[HTML]{D9EAD3}{mismatched}.}
\label{tab:hybrid_workload_result_by_enumerator}
\vspace{-2mm}
\scriptsize
\begin{tabular}{p{1.1cm}rrrrr}
\toprule
\multirow{2}{*}{Enumerator} & \multirow{2}{*}{Eval. workload} & \multicolumn{4}{c}{Construction workload} \\
\cmidrule{3-6}
& & \job{} & \scale{} & Shrink & \job{}+\scale{} \\
\midrule
\multirow{2}{*}{\uniview{}}
& \job{}   & \cellcolor[HTML]{E4DFEC}{20.02} & \cellcolor[HTML]{EBF1DE}{6.30} & 3.18 & 23.69 \\
& \scale{} & \cellcolor[HTML]{EBF1DE}{0.00}  & \cellcolor[HTML]{E4DFEC}{13.32} & $+\infty$ & 6.39 \\
\midrule
\multirow{2}{*}{\ecse{}}
& \job{}   & \cellcolor[HTML]{E4DFEC}{49.21} & \cellcolor[HTML]{EBF1DE}{35.63} & 1.38 & 46.63 \\
& \scale{} & \cellcolor[HTML]{EBF1DE}{13.47} & \cellcolor[HTML]{E4DFEC}{9.86}  & 0.73 & 10.73 \\
\bottomrule
\end{tabular}
\end{table}

Two patterns stand out. First, \ecse{} is substantially more robust to workload drift than \uniview{}: its shrink factors stay near 1 (1.38 and 0.73), whereas \uniview{} degrades sharply (3.18 and $+\infty$). The likely reason is that \ecse{} derives candidates from workload-level join graphs, which can transfer across related query mixes, while \uniview{} produces more predicate-specialized candidates that may fail to match shifted workloads. Second, cross-workload construction can sometimes outperform in-distribution construction: for \ecse{} on \scale{} evaluation, constructing from \job{} yields 13.47\% time saving, higher than 9.86\% from \scale{} construction. This suggests that a richer historical workload can expose reusable join structures that benefit a simpler future workload.

\subsubsection{Data-distribution skew and hardware resource pressure.}\label{sec:data_skewness_short}
The other two robustness checks affect different stages. Under \dsb{} skew, realized time saving and native rewrite success both vary substantially across engines, indicating that skew affects not only original-query runtime but also optimizer-level MV usage. Under memory pressure, selected MV sets often remain useful across environments, but native rewrite success can still change, reinforcing that rewriting remains engine-sensitive. Detailed evidence is reported in appendix (\textsection \ref{sec:robustness-data}).
\section{Case Study}
\label{sec:exp-case}

\begin{figure}
    \centering
    \includegraphics[width=1.1\linewidth]{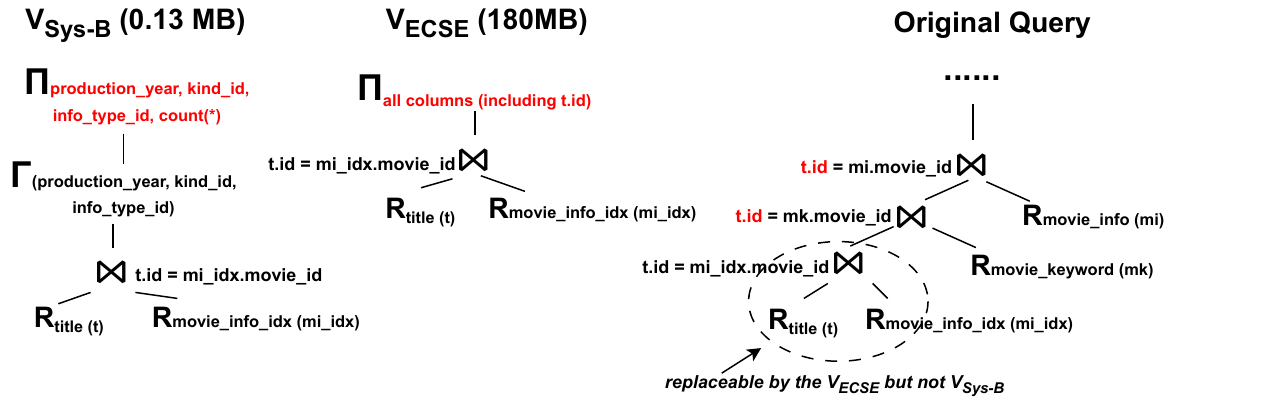}
    \vspace{-8mm}
    \caption{Rewriting Failure Case of \celerdata}
    \label{fig:case-rewrite-failure-sysb}
\end{figure}

\subsection{Case 1: Comparing Non-commercial vs. \celerdata{} Enumerator Across Workloads}
\label{section:case_study_on_enum}

From the results in Figure \ref{fig:pg_detailed_result} and Figure \ref{fig:exp-cross-engine}, we observe that \celerdata{} exhibits large time savings and high query coverage on \tpcds{} as an enumerator. A natural question is why these benefits are drastic on \tpcds{} but do not consistently transfer to the other workloads.
A key design choice of \celerdata{} is aggressive column pruning when generating view candidates, producing substantially lighter weight materialized views (Section~\ref{sec:enumeration} and Figure~\ref{fig:enumerator_different_behaviour}). This strategy is particularly effective on \tpcds{}. For example, the 25 views enumerated and recommended by \celerdata{} occupy only 30~MB of storage space in total, whereas a comparable view generated by \ecse{} --- sharing the same join pattern but without column pruning --- takes 2.3~GB. Under a space budget of up to 1 GB, \celerdata{} therefore provides many more feasible candidates.

However, the same column pruning can also lead to more rewrite failures and, consequently, weaker performance on other workloads than column-pruning-free enumerators. To illustrate, Figure~\ref{fig:case-sysb-enumerator} (in appendix) and Figure~\ref{fig:case-rewrite-failure-sysb} show an original query from workload \scale{} and two selected views: one enumerated by \ecse{} ($V_{\ecse}$) and the other by \celerdata{} ($V_{\celerdata}$). While the two views share the same join over two relations, $V_{\ecse{}}$ retains all columns, whereas $V_{\celerdata}$ keeps only three and omits \texttt{title.id}. In the original query, \texttt{title.id} is required for joining additional relations; as a result, only $V_{\ecse{}}$ can be used for rewriting. 

This explains why \celerdata{} appears particularly strong on \tpcds{}. Its database is substantially larger, so a fixed 1~GB budget is relatively tight compared to the database scale. Under such constraints, pruning-free enumerators are more likely to generate candidates that exceed the budget, even if they provide better coverage in general. In contrast, \celerdata{}'s pruning yields many budget-feasible candidates and thus performs well. For the other workloads with smaller scale, the same 1~GB budget is effectively more generous: many pruning-free candidates remain feasible, and the dominant limitation shifts from storage to query coverage. In short, no single enumerator dominates across all workloads, which motivates evaluating multiple enumerators under different budget settings.

\subsection{Case 2: Diagnosing Rewrite Gaps Between \hive~ and \doris}
\label{sec:case-rewriter-summary}
We diagnose the rewrite gap between \hive~ (3,200 pairs) and \doris~ (747 pairs) on \stats+\ecse, which shows two structural advantages of \hive. First, join-matching flexibility: \doris~ rewrites a view only when the join graph matches a small set of pre-recognized patterns; additional joined relation breaks the match. Second, predicate pushdown: when a query carries additional selection predicates beyond the view definition, \hive~ applies them atop the view scan, while \doris~ rejects the rewrite due to stricter predicate-compatibility enforcement. More details are elaborated in appendix.

\subsection{Case 3: \bigsubs{} Under-ranks High-Coverage Views (\stats, 102\,MB)}
\label{sec:case-bigsubs-ranking}

\noindent To make the \bigsubs{} ranking-failure mechanism (introduced in Section~\ref{sec:stage_wise_analysis}) concrete, we present the smallest-budget instance of the gap. Setup: workload~\stats, enumerator~\hawc, rewriter~\hive, engine~\pg, and a tight 102\,MB budget. Both \bigsubs{} and \gnnmv{} are pure-ILP selectors that consume the same candidate pool; the question is which views each selects.

\noindent \textbf{Headline gap.} Under this configuration, \bigsubs{} achieves 1.71\,\% workload time saving while \gnnmv{} achieves 44.11\,\% --- a 42.4\,pp gap. Both selectors saturate the budget (\bigsubs{} 23 views at 100\,\% utilization; \gnnmv{} 58 views at 99.97\,\%; Table~\ref{tab:case-bigsubs-budget-utilization}), so the gap \emph{cannot} be explained by budget under-utilization. The disagreement is in \emph{which} views the two selectors pick.
{\scriptsize
\begin{table}[htbp]\centering
\caption{Budget utilization rate and time saving achieved by each selector on \stats~$+$~\hawc~at the 102\,MB budget. Rewriter = \hive, engine = \pg.}
\label{tab:case-bigsubs-budget-utilization}
\vspace{-2mm}
\begin{tabular}{lrrr}\toprule
\textbf{Selector} & \textbf{MV count} & \textbf{Utilization (\%)} & \textbf{Time saving (\%)} \\\midrule
\bigsubs & 23 & 100.00 & 1.71 \\
\gnnmv   & 58 & 99.97 & 44.11 \\
\uniview &  1 &  3.43 & 0.02 \\
\bottomrule
\end{tabular}
\end{table}
}

\noindent \textbf{Where the rankings disagree.} Table~\ref{tab:case-bigsubs-ranking} shows the three highest-utility views chosen by \bigsubs{} (top half) and by \gnnmv{} (bottom half). Two patterns are immediate. First, \bigsubs{}'s top three views (MV~121, 263, 7340) collectively contribute essentially zero workload time saving --- 0.00, 0.03, 0.00\,\% respectively --- yet \bigsubs{} ranks them highest because their utility numbers (4.64, 4.15, 3.03) are the largest in its model. Second, the three views that account for the bulk of \gnnmv{}'s 44.11\,\% time saving (MV~1017, 3232, 906; actual contributed 8.94, 19.26, 12.55\,\%) appear far down \bigsubs{}'s utility ranking (estimated benefit 2.9, 1.35, 1.04). \bigsubs{} \emph{has these views in its candidate pool} but under-ranks them out of the budget.

{\scriptsize
\begin{table}[h]\centering
\caption{Top-three views ranked by \bigsubs{} versus top-three ranked by \gnnmv{} on \stats~$+$~\hawc~$+$~102\,MB. \emph{Estimated benefit} is each selector's utility score; \emph{actual contributed time saving} is measured post-rewrite on \pg.}
\label{tab:case-bigsubs-ranking}
\vspace{-2mm}
\begin{tabular}{p{0.7cm}p{0.9cm}p{0.9cm}p{0.7cm}p{1.2cm}p{1.2cm}}\toprule
\textbf{MV idx} & \textbf{Est.\ benefit (\bigsubs)} & \textbf{Est.\ benefit (\gnnmv)} & \textbf{Mat.\ size (MB)} & \textbf{Eventually selected by} & \textbf{Actual contributed TS (\%)} \\\midrule
121  & 4.64 & 6.15   & 44.50 & \multirow{3}{*}{Only \bigsubs} & 0.00 \\
263  & 4.15 & 0.35   & 41.00 &                                & 0.03 \\
7340 & 3.03 & 0      & 10.00 &                                & 0.00 \\\midrule
1017 & 2.9  & 73.05  & 35.50 & \multirow{3}{*}{Only \gnnmv}   & 8.94 \\
3232 & 1.35 & 218.77 & 36.50 &                                & 19.26 \\
906  & 1.04 & 40.44  & 10.50 &                                & 12.55 \\
\bottomrule
\end{tabular}
\end{table}
}

\noindent \textbf{Mechanism.} \bigsubs{}'s utility is computed as $u = (\textit{creation\_cost} - \textit{scan\_cost}) \times \textit{count}$. Table~\ref{tab:case-bigsubs-utility-breakdown} decomposes this for the relevant views: \bigsubs{}'s top pick MV~7340 has high per-view utility ($u = 3.03$) because its creation cost ($3.11$) is large, but it covers only 1 query. \gnnmv{}'s actually helpful picks (MV~1017, 3232, 906) cover many queries (39, 12, 28) but their per-view utility ($a-b = 0.07, 0.11, 0.04$) is tiny because their creation cost is modest. Multiplying by \textit{count} cannot lift them above \bigsubs{}'s creation-cost-dominated picks, so the moderate-cost / high-coverage views never enter \bigsubs{}'s top set. This is the same mechanism that explains the 47.53\,pp \bigsubs{}/\gnnmv{} gap on JOB+1\,GB reported in Section~\ref{sec:exp-endtoend}; the STATS \\+ 102\,MB instance shown here is simply the most extreme case in our budget sweep. The takeaway is that \bigsubs{}-style cost-only utility models systematically penalize the precise heavy-tail views (moderate creation cost, high query coverage) that \ecse-class enumeration is best at producing.

{\scriptsize
\begin{table}[h]\centering
\caption{\bigsubs{} utility breakdown for the relevant views on \stats~$+$~\hawc~$+$~102\,MB. Recall $u = (a - b) \times \textit{count}$, where $a$ is creation cost and $b$ is sequential scanning cost.}
\label{tab:case-bigsubs-utility-breakdown}
\vspace{-2mm}
\begin{tabular}{lp{1.0cm}p{1.0cm}p{0.7cm}p{1.0cm}p{1.0cm}}\toprule
\textbf{MV idx} & \textbf{Creation cost ($a$)} & \textbf{Scan cost ($b$)} & \textbf{Per-view utility ($a-b$)} & \textbf{Covered query count} & \textbf{Estimated utility ($cnt \cdot (a-b)$)} \\\midrule
7340 & 3.11 & 0.08 & 3.03 & 1  & 3.03 \\\midrule
3232 & 0.40 & 0.29 & 0.11 & 12 & 1.35 \\
1017 & 0.36 & 0.29 & 0.07 & 39 & 2.90 \\
906  & 0.12 & 0.08 & 0.04 & 28 & 1.04 \\
\bottomrule
\end{tabular}
\end{table}
}
\section{Discussion}
\label{sec:discussion}
Our evaluation shows that MV-based rewriting remains workload- and engine-dependent: end-to-end gains are governed by interactions among candidate enumeration, budget-constrained selection, and rewrite applicability. We therefore summarize the main diagnostic signals for practitioners in Table~\ref{tab:practical-guidance}; the goal is not to prescribe a universal pipeline, but to identify which stage is likely to be limiting under observable workload and engine properties.

\noindent\textbf{Practical Guidance} Drawing only on evidence already presented above, we summarize the cross-stage findings in a compact decision table (Table~\ref{tab:practical-guidance}) and a metrics-to-watch checklist for practitioners building MV-based query-rewriting pipelines.

\begin{table}[h]\centering\small
\caption{Workload-feature $\to$ method-choice decision suggestion, with paper-side evidence pointers. 
}
\label{tab:practical-guidance}
\vspace{-2mm}
\scriptsize
\begin{tabular}{p{1.5cm}p{1.9cm}p{2cm}p{1.6cm}}\toprule
\textbf{Workload feature} & \textbf{Symptom / metric} & \textbf{Recommended action} & \textbf{Evidence} \\\midrule
Join-heavy queries; heavy-tailed view coverage & Top-1 view coverage substantially above the median (e.g., $>$30\,\% on \job{}/\stats{}) & Use cross-query-pattern enumerators such as \ecse{}; plan-confined enumerators may miss high-leverage views & Table~\ref{tab:enum_candidate_stats}; Fig.~\ref{fig:query_coverage_vs_size} \\\midrule
Tight budget vs.\ large base data & Budget-to-DB ratio is small; useful candidates exceed budget & Prefer compact or column-pruned candidates; inspect coverage-vs-size trade-off  & Case study \textsection\ref{section:case_study_on_enum}; Table~\ref{tab:top_one_combination_of_each_workload_engine_scenario} \\\midrule
High predicate / type-conversion diversity & Predicate structure is diverse (\stats-like profile)  & Validate selector ranking carefully; learned selectors such as \gnnmv{} may help.  & Table~\ref{tab:workload_selection_predicate_operator-stat}; selector analysis in \textsection\ref{sec:stage_wise_analysis} 
 \\\midrule
Plan-transparent engine, or low rewrite-success & Native rewrite-success is much lower than portable \hive{} on the same MV set & Validate with a portable \hive{} rewrite baseline; inspect the negative-speedup tail before deploying. & Table~\ref{tab:exp-rewrite-success}; Fig.~\ref{fig:exp-cross-engine}; 
Table~\ref{tab:skewed_distribution_rewritability} \\

\bottomrule
\end{tabular}
\end{table}

\noindent\textbf{Key metrics.} Users should monitor three classes of diagnostics: candidate reuse potential, budget sensitivity and rewrite risk. Candidate reuse (enumeration) is captured by view coverage and coverage-vs-size; budget sensitivity (selection) is captured by time-saving curves under different storage budgets; and rewrite risk (rewriting) is captured by MV usage rate and the negative-speedup tail. 

\noindent\textbf{Limitations and Future Work} First, cross-engine comparison cannot fully eliminate system-specific effects. Even though our protocol has controlled multiple factors, some optimizer decisions still remain opaque in plan-transparent systems. Our results should therefore be interpreted as a comparison of practical MV-exploitation behavior, not as a fully isolated comparison of rewrite algorithms.
Second, we evaluate MVs statically. We do not model MV refresh cost, update frequency, or maintenance overhead. Incorporating maintenance-aware objectives and richer optimizer observability is important directions for extending this benchmark. We leave these valuable topics for future work.

\section{Conclusion}
\label{sec:conclusion}

We presented an end-to-end, cross-engine evaluation of MV-based query rewriting across candidate enumeration, view selection, and query rewriting. We show pipeline components cannot be ranked independently: a method that appears best under one fixed context may not be best once the candidate space, selector, rewriter, and engine change together. Making MV-based rewriting predictable in practice will therefore require co-designing high-leverage candidate generation, budget-robust selection, and rewrite applicability with explicit regression control. Our modular framework and cross-engine protocol provide a basis for reproducible comparison.

\begin{acks}
We acknowledge the support of the Natural Sciences and Engineering Research Council of Canada (NSERC).
\end{acks}
\bibliographystyle{ACM-Reference-Format}
\bibliography{main}
\clearpage
\appendix
\section{Extended Related Work}
\label{app:related-work}

\subsection{View Enumeration}
\label{app:enum-related}
Early work in data warehouses considered enumerating candidates by exploring data cubes. These works~\cite{sigmod_1996_Harinarayan_stanford,ho1997range} model group-by aggregates as nodes in a data cube lattice and consider all lattice nodes as view candidates, capturing the dependency between ancestry and descendants in the lattice. Later works~\cite{gupta1997selection, Gupta2005Selection} generalize the notion of lattice to graphs that capture the relationships between views, i.e., whether a view can be computed from other views or tables.
Since the space of the lattice and the generalized graphs is huge, researchers propose methods from the perspective of exploring the query workload. Agrawal et al.~\cite{vldb_2000_automated_mv_selection_microsoft} proposed to find the syntactical sub-structures (single-block materialized views consisting of selection, join, grouping, and aggregation, which use a subset of tables from a query in the workload) as candidate views and then prune and merge the views based on a query optimizer. Aouiche et al.~\cite{Aouiche2006Clustering-based} extend this method by clustering queries by similarity (using heuristics like overlapping attributes, tables, predicates) to capture closely related queries.
Another line of work~\cite{Zhou2007Efficient,giannikis2014shared,vldb_2018_bigsubs_microsoft,Jindal2018Selecting,Jindal2018Computation} shares a similar idea, originated from multi-query optimization, by looking into the query workload and finding the common subexpressions (e.g., SPJ+GroupBy in~\cite{sigmod_2001_optimizing_using_mv_microsoft}) in queries as the source of candidates. The identified common subexpressions can be helpful for either temporary reuse or materialization. Each of the subexpression-based works may propose techniques for cheaply detecting potentially sharable subexpressions and pruning the enumeration space. For example,~\cite{Zhou2007Efficient} proposed the table signatures as a fast filter to prune non-sharable expressions, and BigSubs~\cite{vldb_2018_bigsubs_microsoft} filters out infrequent subexpressions (subtrees in the logical query plan).
The four representative methods evaluated in Section~\ref{sec:enumeration} (Table~\ref{tab:enum_mapping}) instantiate, respectively, the plan-subtree, join-graph/workload, MVPP/learned-style, and integrated-industrial categories of this design space.

\subsection{View Selection}
\label{app:select-related}
View selection chooses a budget-constrained subset of candidates to materialize. While some works target the storage-budget-unconstrained scenario~\cite{sigmod_2000_roy_iit,icdcs_1997_yang_unsw,dexa_1998_Ligoudistianos_athens,vldb_1997_Theodoratos_athens,soft_computing_2003_Horng,dawak_1999_zhang,edbt_2009_chaves_sap,dba_2006_Derakhshan_Griffith,ICA3PP_2008_Derakhshan_Griffith,cyber_2001_zhang_unsw}, most consider the budgeted setting, which is also ours. An early branch uses traditional heuristic search over candidates modeled as Cube Lattices~\cite{waim_2005_ye,ssdbm_2003_Bauer_gmbh,sigmod_1996_Harinarayan_stanford,ieee_2003_Jeffrey,data_know_2002_Kalnis_hkust} or AND/OR view graphs~\cite{caise_2003_baril_Montpellier,gupta1997selection,inf_sys_2001_lee,DEXA_2011_mami_Montpellier}, with greedy~\cite{sigmod_2001_hoshi_iit,Gupta2005Selection}, dynamic-programming~\cite{data_eng_conference_2014_perez_rice}, or integer-programming~\cite{vldb_2018_bigsubs_microsoft} cost-based pruning. Another early branch resorts to genetic algorithms over the \mvpp~\cite{vldb_1997_mvpp_unsw} to navigate spaces where heuristics stagnate. More recent learned approaches---DQM~\cite{arxiv_2019_dqm_chicago}, AutoView~\cite{icde_2021_autoview_thu}, and UniView~\cite{vldb_2024_uniview_zju}---use neural networks to learn query/view representations and a Q-learning RL framework to estimate view benefit and select views. Beyond academia, several commercial systems (e.g., \celerdata) couple enumeration and selection into a single integrated, black-box recommender.

\noindent\textbf{\bigsubs{} ILP formulation.} \bigsubs~\cite{vldb_2018_bigsubs_microsoft} formulates selection as an integer linear program maximizing the total utility of selected views. The utility of a view $v$ for a query $q$ is the execution-cost reduction from rewriting $q$ with $v$, i.e., $\cost(q) - \cost(q \mid v)$ (with multiple possible rewrites, only the maximum reduction for $q$ is counted). It further models view interactions by accounting for query-plan overlap (one view being a subtree of another), discouraging selecting both overlapping views, and the ILP enforces the storage-budget constraint directly. This cost-only utility is the source of the under-ranking behavior analyzed in \textsection \ref{sec:stage_wise_analysis} and \textsection \ref{sec:case-bigsubs-ranking}.

\noindent\textbf{Our implementation.} Most academic and industrial selectors are closed-source. We implemented the non-learning \bigsubs{} (ILP formulation) and the learning-based \gnnmv; for \uniview{} we adopted its publicly available RL-based implementation (similar in framework to \autoview). For both learning-based methods we followed the offline training process in the original papers and trained models separately per dataset.

\subsection{Query Rewriting}
\label{app:rewrite-related}
Using views in query answering has been widely applied in query optimization in data warehouses~\cite{vldb_2001_Halevy_washington,sigmod_1996_Harinarayan_stanford} and data integration systems~\cite{vldb_1996_levy_at&t,sac_1997_Duschka_stanford,IJCAI_1999_eric,vldb_1999_Daniela_oracle}. Early works developed query rewriting algorithms based on query containment and equivalence on conjunctive queries~\cite{stoc_1977_Chandra_ibm,acm_1981_sagiv_princeton,acm_1988_klug_madison,vldb_1993_levy_stanford,dood_1993_zhang_intel,pods_1998_Kolaitis_santacruz}, but the queries handled then were too simple for real-world cases. Later works extended these foundations to richer query classes---grouping and aggregation~\cite{pods_1999_Grumbach_inria,pods_1999_cohen_jerusalem,vldb_1995_gupta_ibm,afrati2005selecting,icdt_1999_gupta_stanford}, comparison conditions~\cite{pods_1995_levy_at&t,pods_2002_foto_athens,Cohen2006Rewriting}, and set operations~\cite{acm_1981_sagiv_princeton}---which often lead to higher complexity or undecidability~\cite{Abiteboul1995Foundations,vldb_2001_Halevy_washington}.
In terms of implementation, evolving from early pure rule-based approaches~\cite{gupta1995aggregate,sigmod_2001_optimizing_using_mv_microsoft,park2001rewriting}, the prevailing architecture today integrates the rules into a System-R~\cite{systemr}-style query optimizer~\cite{vldb_1998_bello_oracle,icde_1995_Chaudhuri_hplab,vldb_1996_Tsatalos_ibm}, where view-based rewrite is part of the plan search space and cooperates with cost estimation when forming the final execution plan.

\section{Workload Characterisitcs} \label{sec:workload_char_supp}

{
\scriptsize
\begin{table}[H]\centering
\caption{Query Count and Join count Summary Statistic of Each Workload}\label{tab:workload_query_cnt_join_cnt}
\vspace{-4mm}
\begin{tabular}{lC{0.5cm}C{0.5cm}C{0.8cm}C{0.6cm}|C{0.4cm}C{0.4cm}C{0.7cm}}\toprule
\multirow{2}{*}{\textbf{Workload Name}} &\multicolumn{4}{C{2.4cm}}{\textbf{Query Count}} &\multicolumn{3}{C{1.3cm}}{\textbf{Join Count}} \\\cmidrule{2-8}
&\textbf{Total} &\textbf{Train} &\textbf{Validation} &\textbf{Test} &\textbf{min} &\textbf{max} &\textbf{median} \\\midrule
JOB &113 &33 &11 &69 &3 &16 &7 \\
SCALE &500 &150 &50 &300 &0 &4 &2 \\
STATS &1,449 &434 &144 &871 &1 &6 &3 \\
TPCDS &938 &281 &93 &564 &2 &7 &3 \\
\bottomrule
\end{tabular}
\end{table}
}

{
\scriptsize
\begin{table}[H]\centering
\caption{Average occurrence frequency of each selection-predicate operator per query, by category. \revc{\texttt{WHERE} operator} counts are extracted from the logical query plan parsed by sqlglot.}\label{tab:workload_selection_predicate_operator-stat}
    \vspace{-3mm}
\begin{tabular}{C{2.2cm}C{1.7cm}C{0.5cm}C{0.6cm}C{0.6cm}C{0.8cm}}\toprule
\textbf{Category} &\textbf{Operators} &\textbf{JOB} &\textbf{SCALE} &\textbf{STATS} &\textbf{TPCDS} \\\midrule
\textbf{General Arithmetic} &+, -, Negation &0 &0 &1.00 &\textbf{1.96} \\
\textbf{Comparison} &=, !=, <, >, <=, >= &3.57 &2.10 &\textbf{6.20} &4.20 \\
\textbf{Logical} &And, Or, Not &\textbf{6.30} &1.85 &5.44 &3.89 \\
\textbf{Null Judgement} &Is Null &\textbf{1.17} &0 &0 &0 \\
\textbf{Range} &In, Between &1.90 &0 &0 &\textbf{2.67} \\
\textbf{String Operation} &Substring, Like &\textbf{2.40} &0 &0 &1.43 \\
\textbf{Time Operation} &Interval &0 &0 &0 &1.17 \\
\textbf{Type Conversion} &Cast, DataType &0 &0 &\textbf{5.79} &4.00 \\
\bottomrule
\end{tabular}
    \vspace{-3mm}
\end{table}
}
{
\footnotesize
\begin{table}[t]
  \centering
  \caption{Adoption timeline of MV capabilities in production systems (year = first public documentation; ``--'' =  not exposed or not supported).}
  \vspace{-3mm}
  \begin{tabular}{lccc}
    \toprule
 System & MV support & Auto MV selection & Auto MV rewrite \\
\midrule
BigQuery~\cite{bigquery} & 2021 & 2024 & 2021 \\
Snowflake~\cite{snowflake} & 2018 & -- & 2018 \\
Redshift~\cite{redshift} & 2019 & 2021 & 2020 \\
Oracle  \cite{oracle}& 1998  & 2021 & 2009 \\
Doris~\cite{doris} & 2024 & -- & 2024 \\
StarRocks~\cite{starrocks} & 2022 & -- & 2023 \\
Hive \cite{hive}&2017 & --  & 2018 \\
Celerdata \cite{celerdata}&2022 & 2022  & 2023 \\

PostgreSQL \cite{pg}& 2013 & -- & -- \\
    \bottomrule
  \end{tabular}
  \label{tab:mv-rewrite-systems}
  \vspace{-4mm}
\end{table}

}

\section{Detailed Comparison on Rewriters}

We now compare these query rewriters conceptually in this subsection. Unlike view enumeration and view selection, which are largely evaluated through academic prototypes, query rewriting is tightly coupled with query optimizers. We therefore focus on full-fledged systems that support view-based rewriting for general SQL workloads.  As a result, these systems differ in multiple dimensions beyond algorithmic design, including optimizer integration, user visibility, coverage of query classes, and scope of rewrite rules.

Table~\ref{tab:rewriter-comparison-conceptual} provides a high-level comparison based on our examination of the official documentation of Apache Calcite and the evaluated data warehouse systems, as well as their codebases and GitHub issues (if open-sourced).

{
\scriptsize
\begin{table}[!htp]\centering
\caption{Comparison of selected rewriters on key dimensions. 
SPJA = plan-level transformation rules for relational operators selection, projection, join, and aggregation.
}\label{tab:rewriter-comparison-conceptual}
\vspace{-4mm}
\begin{tabular}{p{0.8cm}p{1.2cm}p{0.7cm}p{1.6cm}p{2.4cm}}\toprule
\textbf{Rewriter} &\textbf{Rewrite level} &\textbf{Cost-aware?} &\textbf{Pre-rewrite plan normalization} &\textbf{Rewrite rule composition} \\\midrule
\textbf{\calcite} & SQL \& logical plan & No &No &SPJA, no join reordering \\
\textbf{\hive} &SQL \& logical plan & Yes & Yes &SPJA + predicate compensation \\
\textbf{\doris} & Physical plan & Yes &No &SPJA + predicate compensation + extended operators like window function \\
\textbf{\starrocks} &Physical plan& Yes &No & SPJA + predicate compensation + AST-level text matching \\
\textbf{\redshift} &Physical plan & Unknown &Unknown &Unknown \\
\textbf{\celerdata} &Physical plan & Unknown &Unknown &Unknown \\
\textbf{\bigquery} &Physical plan & Unknown &Unknown &Unknown \\
\bottomrule
\end{tabular}
\end{table}
}

\noindent\textbf{Cost-aware rewrite:} All selected approaches adopt a System-R-style framework~\cite{systemr}, where view-based query rewrite is implemented as a transformation integrated into the query optimizer. The vanilla \calcite~ slightly deviates from this paradigm: it applies rewrites without cost-based decision making, whereas the other approaches explicitly compare plan costs to determine whether a plan involving materialized views should be chosen.

\noindent\textbf{Overall rewriting workflow:} All evaluated rewriters follow a rule-based subexpression matching and substitution paradigm: subexpressions in the logical plan are matched against available materialized views and replaced when a match is found. 
When a view is more general than the matched subexpression (e.g., it omits some selection predicates), the rewriter may perform predicate compensation: additional selection predicates that are present in the original query but missing from the view definition are derived and applied on top of the view scan to guarantee an equivalent rewrite.
The systems differ in the preprocessing steps before rule-based rewriting. \hive~ performs a series of normalizations prior to rewrite, transforming expressions into canonical forms (e.g., expanding \texttt{IN} and \texttt{BETWEEN} predicates), followed by further plan transformations such as predicate pull-up to increase rewrite opportunities. 
In contrast, \calcite, \starrocks, and \doris~ perform fewer such preprocessing steps and apply rule-based matching and replacement directly to the original logical plan.

{
\scriptsize
\begin{table}[!htp]\centering
\caption{Classification of predicate types. Source: Code base of \starrocks. \href{https://github.com/StarRocks/starrocks/blob/main/fe/fe-core/src/main/java/com/starrocks/sql/optimizer/rule/transformation/materialization/PredicateSplit.java}{link}.}\label{tab:rewritter-predicate-types}
\vspace{-4mm}
\begin{tabular}{lp{3cm}p{3cm}r}\toprule
\textbf{Predicate Type} &\textbf{Definition} &\textbf{Example Form} \\\midrule
1. \textit{equal} &column equality predicates conjuncts &Ti.Cp =Tj.Cq \\ \midrule
2. \textit{range} &range predicates conjuncts &Ti.Cp \textbf{op} $c$, where $c$ is a constant and \textbf{op} is one of the operators: $<, \leq, =, \geq, >$ \\\midrule
3. \textit{residual} &conjuncts can be pulled up but do not belong to \textit{equal} and \textit{range} &Ti.Cp \texttt{LIKE} "\%abc\%" 

\\\midrule
4. \textit{not-pulled-up} &conjuncts cannot be pulled up and should match exactly between query and view &window function \\
\bottomrule
\end{tabular}
\end{table}
}

\noindent\textbf{Supported predicate types in compensation:}  Predicate compensation plays an important role in the success rate of view-based query rewrites. Predicates can be classified into four categories, as summarized in Table~\ref{tab:rewritter-predicate-types}, with the first three types commonly appearing in analytical query workloads. 
In our evaluation, \emph{Vanilla Calcite}, \calcite, does not perform predicate compensation;
all evaluated rewriters support compensation for \emph{equality} and \emph{range} predicates. \starrocks~ and \doris~ additionally provide limited support for \emph{residual} predicates, while only \doris~ supports compensation for \emph{not-pulled-up} predicates. Overall, \doris~ supports a broader range of less common predicate patterns.

\noindent\textbf{Rule composition strategies:} All evaluated rewriters support rule-based view replacement for core relational operators, including selection, projection, join, and aggregation. Predicate compensation is supported by most systems, but not by our \calcite~ baseline, which uses only Calcite’s \texttt{SubstitutionVisitor} with default view-related rules. For join-related rewrites, the DBMS-integrated rewriters (e.g., Hive, StarRocks, and Doris) can match join subexpressions up to join reordering by exploiting join commutativity and associativity. Besides these standard capabilities, Doris additionally supports window joins, while StarRocks provides an alternative rewrite path, \emph{text-match-based} rewriting, which performs view matching and replacement directly at the query’s abstract syntax tree (AST) level. This AST-level path bypasses logical plan construction and thus avoids subsequent steps such as predicate compensation. However, this alternative approach is restrictive, as it requires the view definition to differ only minimally from the original query.

{
\footnotesize
\begin{table}[!htp]\centering
\caption{Qualitative comparison on rewriters. }\label{tab:qualatative_comparison_rewriter}
\begin{tabular}{lp{1.4cm}p{1.7cm}p{3.1cm}}\toprule
Rewriter &Existing Form &Cross Executable &Level of View Definition Requirement \\\midrule
\calcite &Separate &\ding{51} &Logical Registration \\
\hive &Embedded &\ding{51} &Logical Registration\\
\doris &Embedded &\ding{55} &Physical Materialization \\
\starrocks &Embedded &\ding{55} &Physical Materialization \\
\redshift &Embedded &\ding{55} &Physical Materialization \\
\celerdata &Embedded &\ding{55} &Physical Materialization \\
\bigquery &Embedded &\ding{55} &Physical Materialization \\
\bottomrule
\end{tabular}
\end{table}
}
As illustrated in Table \ref{tab:qualatative_comparison_rewriter}, we qualitatively compare these approaches from the following three aspects.

\noindent \textbf{Level of Definition Required.} It indicates in order to successfully carry out query rewriting, whether a rewriter requires the resulting tuples be populated to the view table (physical materialization) or not (logical registration). Although we execute \texttt{CREATE MATERIALIZED VIEW} in either case, the latter one indicates registering the view to the schema conceptually is enough. We measure this dimension by comparing the count of rewritten queries achieved by each rewriter with and without populating tuples. We find that \calcite~ and \hive~ return equal count while the rest of approaches return much fewer without populating tuples. Using logical registration level approaches can save time because there is much fewer I/O and computation overhead when creating the MV. Moreover, as we will show in empirical analysis, such approach can still achieve better performance even with the drawback of failing to exploit precise view cardinalities.
\section{All Combinations Time Saving Result}

\begin{figure*}[htp]\captionsetup[subfigure]{font=footnotesize}
\centering
\caption{Time Saving (\%) Main Result}
\begin{subfigure}[b]{\linewidth}
\centering
\caption{\job~}
    \centering
    \includegraphics[width=\linewidth]{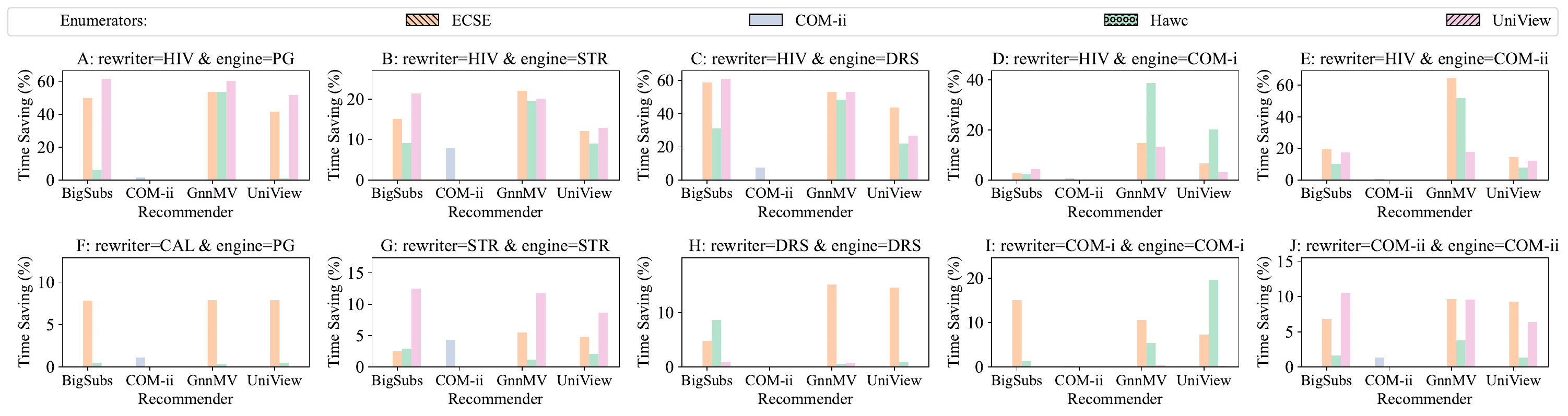}
    \label{fig:ts_main_result_job}
\end{subfigure}
\hfill
\vspace{-10mm}

\begin{subfigure}[b]{\linewidth}
\centering
\caption{\scale~}
    \centering
    \includegraphics[width=\linewidth]{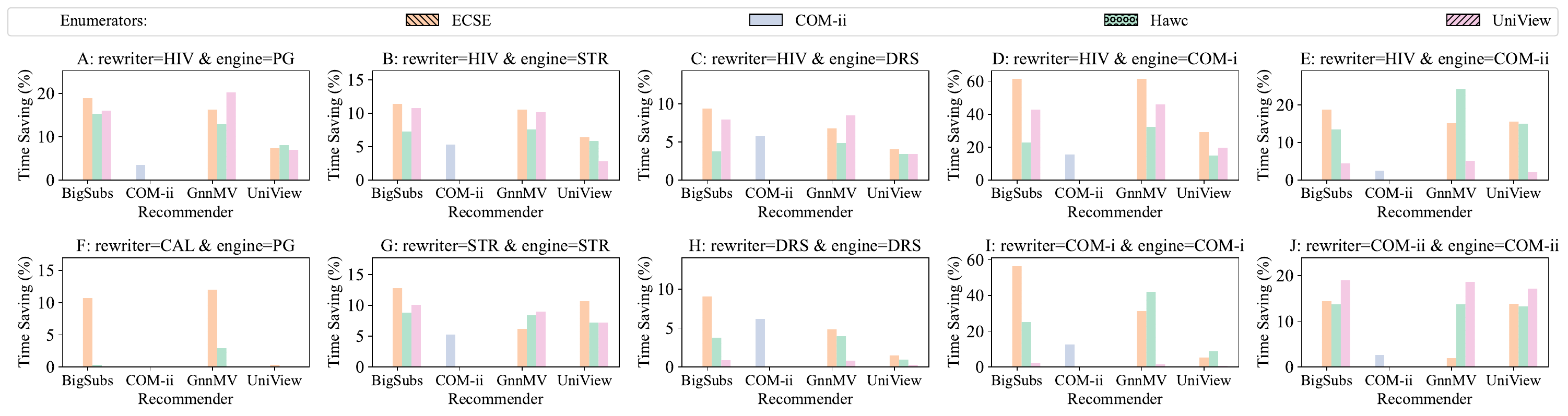}
    \label{fig:ts_main_result_scale}
\end{subfigure}
\hfill
\vspace{-10mm}

\begin{subfigure}[b]{\linewidth}
\centering
\caption{\stats~}
    \centering
    \includegraphics[width=\linewidth]{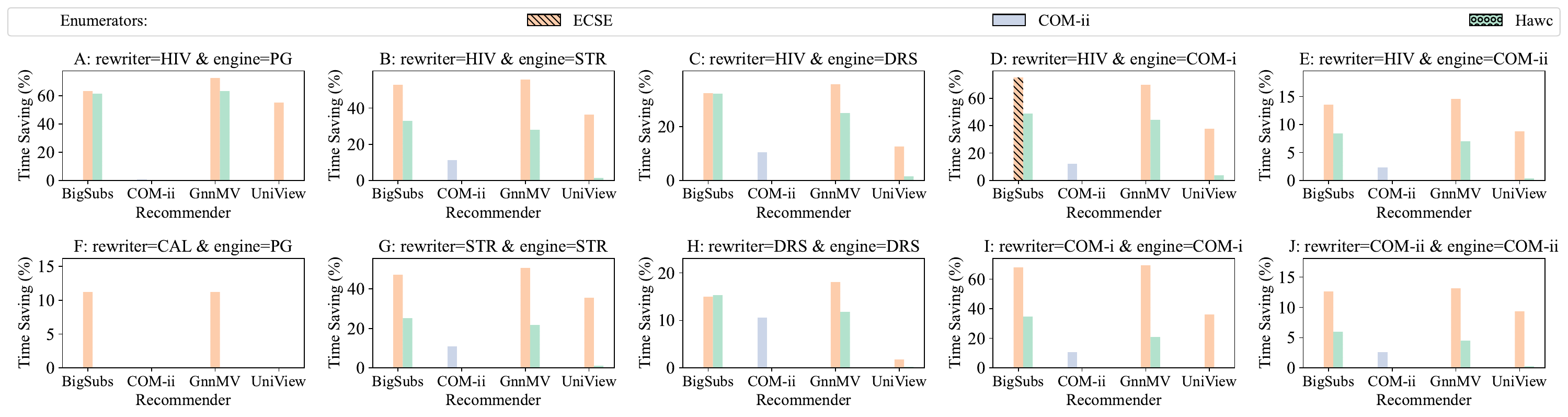}
    \label{fig:ts_main_result_stats}
\end{subfigure}
\hfill
\vspace{-10mm}

\begin{subfigure}[b]{\linewidth}
\centering
\caption{\tpcds~}
    \centering
    \includegraphics[width=\linewidth]{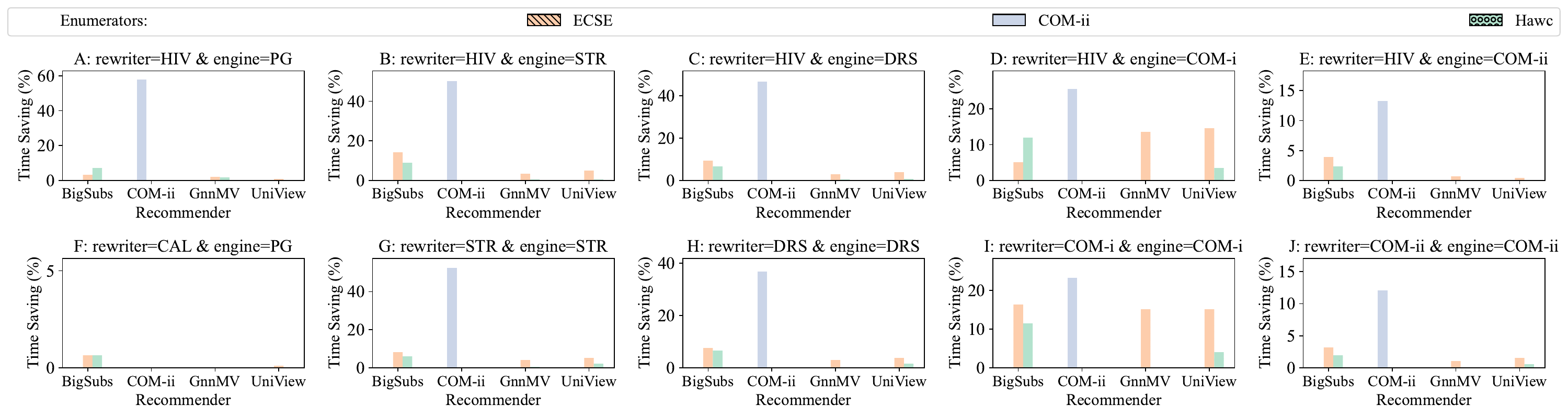}
    \label{fig:ts_main_result_tpcds}
\end{subfigure}

\vspace{-6mm}
\label{fig:ts_main_result}
\end{figure*}

Please refer to Figure \ref{fig:ts_main_result}.

\begin{table}[!htp]\centering
\caption{Top 1 pipeline combination across execution engines and workloads. Storage budget = 1 GB. Median = the median of time savings among all pipeline combinations evaluated.}\label{tab:top_one_combination_of_each_workload_engine_scenario}
{\scriptsize
\begin{tabular}{p{0.8cm}p{0.8cm}|p{0.9cm}p{1.2cm}p{1cm}|cc}\toprule
\multicolumn{2}{c}{\textbf{Scenario Setting}} &\multicolumn{3}{p{3cm}}{\textbf{Top 1 Combination}} &\multicolumn{2}{c}{\textbf{Time Saving (\%)}} \\\cmidrule{1-7}
\textbf{Workload} &\textbf{Engine} &\textbf{Enumerator} &\textbf{Recommender} &\textbf{Rewriter} &\textbf{Top 1} &\textbf{Median} \\\midrule
\multirow{5}{*}{JOB} &DRS &UniView &BigSubs &HIV &61 &15 \\
&PG &UniView &BigSubs &HIV &62 &7 \\
&STR &ECSE &GnnMV &HIV &22 &9 \\
&Sys-A &Basic &GnnMV &HIV &39 &5 \\
&Sys-B &ECSE &GnnMV &HIV &64 &10 \\\midrule
\multirow{5}{*}{SCALE} &DRS &ECSE &BigSubs &HIV &9 &4 \\
&PG &UniView &GnnMV &HIV &20 &7 \\
&STR &ECSE &BigSubs &STR &13 &8 \\
&Sys-A &ECSE &BigSubs &HIV &61 &24 \\
&Sys-B &Basic &GnnMV &HIV &24 &14 \\\midrule
\multirow{5}{*}{STATS} &DRS &ECSE &GnnMV &HIV &36 &14 \\
&PG &ECSE &GnnMV &HIV &73 &6 \\
&STR &ECSE &GnnMV &HIV &55 &30 \\
&Sys-A &ECSE &BigSubs &HIV &75 &37 \\
&Sys-B &ECSE &GnnMV &HIV &15 &8 \\\midrule
\multirow{5}{*}{TPCDS} &DRS &Sys-B &Sys-B &HIV &47 &4 \\
&PG &Sys-B &Sys-B &HIV &58 &1 \\
&STR &Sys-B &Sys-B &STR &52 &5 \\
&Sys-A &Sys-B &Sys-B &HIV &26 &13 \\
&Sys-B &Sys-B &Sys-B &HIV &13 &1 \\
\bottomrule
\end{tabular}
}
\end{table}

\section{Additional Stage-wise Analysis Results}
\label{app:stage-detail}

This appendix holds two secondary stage-wise tables referenced from Section~\ref{sec:stage_wise_analysis}: the structural complexity of enumerated candidates (Table~\ref{tab:enum_complexity}) and the runtime overhead of view selectors (Table~\ref{tab:sys_runtime_recommender}).

{
\begin{table}[t]\centering
\caption{Complexity of candidate views.}\label{tab:enum_complexity}
\vspace{-3mm}
\scriptsize
\begin{tabular}{lrp{1.2cm}p{1.2cm}p{2cm}}\toprule
\textbf{Workload} &\textbf{Enumerator} &\textbf{\# of Unique Join Order} &\textbf{Mean \# of Joins} &\textbf{Mean \# of Unique \texttt{WHERE} Operators} \\\midrule
\multirow{4}{*}{JOB} &\celerdata{} &5 &5.27 &0.64 \\
&\ecse{} &243 &2.34 &0.06 \\
&\hawc{} &78 &2.47 &0.37 \\
&\uniview{} &43 &2.13 &7.98 \\\midrule
\multirow{4}{*}{SCALE} &\celerdata{} &18 &1.30 &0.00 \\
&\ecse{} &30 &1.42 &5.27 \\
&\hawc{} &56 &1.41 &5.41 \\
&\uniview{} &34 &1.58 &7.69 \\\midrule
\multirow{3}{*}{STATS} &\celerdata{} &19 &2.83 &0.00 \\
&\ecse{} &45 &2.44 &5.36 \\
&\hawc{} &32 &2.50 &5.80 \\\midrule
\multirow{3}{*}{TPCDS} &\celerdata{} &18 &3.32 &0.00 \\
&\ecse{} &54 &2.30 &5.99 \\
&\hawc{} &19 &2.44 &7.23 \\
\bottomrule
\end{tabular}
\vspace{-3mm}
\end{table}
}

{
\begin{table}[t]\centering
\caption{Runtime overhead of selectors per workload (sec).}\label{tab:sys_runtime_recommender}
\scriptsize
\vspace{-2mm}
\begin{tabular}{p{2.2cm}rrrrrr}\toprule
\textbf{Phase} &\textbf{Selector} &\textbf{JOB} &\textbf{SCALE} &\textbf{STATS} &\textbf{TPCDS} \\\midrule
\multirow{3}{2cm}{Model Training} &\bigsubs{} & N/A &N/A &N/A &N/A  \\
&\gnnmv{} &287.95 &1495.67 &1551.04 &486.26 \\
&\uniview &2551.00 &6756.21 &13392.20 &6924.05 \\\hline
\multirow{3}{2cm}{Model Inference/ View Selection} &\bigsubs{} &0.26 &0.32 &0.56 &0.40 \\
&\gnnmv{} &3.89 &5.14 &6.06 &4.34 \\
&\uniview{} &20.58 &60.94 &750.84 &45.68 \\
\hline
Enumeration + Selection & \celerdata{} &582.78 &1204.17 &1320.97 &1080.62 \\
\bottomrule
\end{tabular}
\vspace{-3mm}
\end{table}
}

\subsection{Robustness Analysis}
\label{sec:robustness}

\subsubsection{Data-Distribution Drift: \dsb{} vs \tpcds{}}
\label{sec:robustness-data}

\dsb~\cite{dsb} shares the same schema as \tpcds{} but applies a more skewed data distribution to selected fact columns. We use the \tpcds{} instance for enumeration, training, and view selection, then benchmark original and rewritten latencies on both \tpcds~ and \dsb{} instance\footnote{scaling factors are aligned as 10.}. We pick up PostgreSQL as the main engine to be studied and Doris as a reference to be compared with. 

\textbf{Workload preparation.} We split the workload into queries that are sensitive to the distribution drift (\emph{sensitive workload}) and those that are not (\emph{not-sensitive workload}); the split is by relative original-latency change across the two instances in each engine. Table~\ref{tab:skewed_distribution_original_latency_change_rate} reports the per-engine change in original-workload latency when switching from \tpcds{} to \dsb. The not-sensitive workload changes little (\pg{} +8.71\%, \doris~ -3.26\%), while the sensitive workload becomes substantially slower in both engine, most dramatically on \pg{} (+57.65\%).

{
\footnotesize
\begin{table}[!htp]\centering\small
\caption{Original-workload latency change (\%) when the underlying data is switched from \tpcds{} (balanced) to \dsb{} (skewed) for each workload}\label{tab:skewed_distribution_original_latency_change_rate}
\begin{tabular}{lrrrrr}\toprule
\textbf{Workload} &\textbf{Doris} &\textbf{PostgreSQL} \\\midrule
\textbf{not-sensitive} &-3.26\% &8.71\%  \\
\textbf{sensitive} &16.70\% &57.65\% \\
\bottomrule
\end{tabular}
\end{table}
}

\textbf{Data skewness can discount performance but harms more on \gnnmv.} Table \ref{tab:skewed_distribution_ts} shows the average MV-driven workload time saving for the two top selectors (\gnnmv{} and \bigsubs) with \ecse{} as enumerator and \hive{} as rewriter. Unlike Doris, as a more classical cost-based optimizer, PostgreSQL substantially suffers from MV benefits decrease on sensitive workloads when facing instance mismatch. However, we observe although \gnnmv~ enjoys 10\% advantage than \bigsubs~ (71.67\% vs 61.04\%) under matched scenario, it drops over 20\% (71.67\% to 49.71\%) while \bigsubs~ only drops 8\% under mismatch. This is a sign showing the curated optimized training process makes learned model a fine-tuned one perfectly running on designed scenario, but essentially makes it an "overfit" version sticking to one distribution pattern, lacking transferability.

{
\begin{table}[!htp]\centering\small
\caption{Time Saving (\%) on \tpcds{} (balanced) vs \dsb{} (skewed) instances. Storage budgets = 8 GB; enumerator = \ecse, rewriter = \hive.}\label{tab:skewed_distribution_ts}
\scriptsize
\begin{tabular}{lrrrrrr}\toprule
& &\multicolumn{2}{c}{BigSubs recommender} &\multicolumn{2}{c}{GnnMV recommender} \\\cmidrule{3-6}
Workload &Engine &TPCDS ins. &DSB ins. &TPCDS ins. &DSB ins. \\\midrule
sensitive &PostgreSQL &61.04 &53.38 &71.67 &49.71 \\
not-sensitive &PostgreSQL &30.11 &38.68 &29.16 &37.51 \\
sensitive &Doris &10.08 &11.40 &6.45 &8.82 \\
not-sensitive &Doris &49.90 &49.49 &26.42 &22.02 \\
\bottomrule
\end{tabular}
\end{table}
}

\begin{figure}[!htp]
    \centering
    \includegraphics[width=0.7\linewidth]{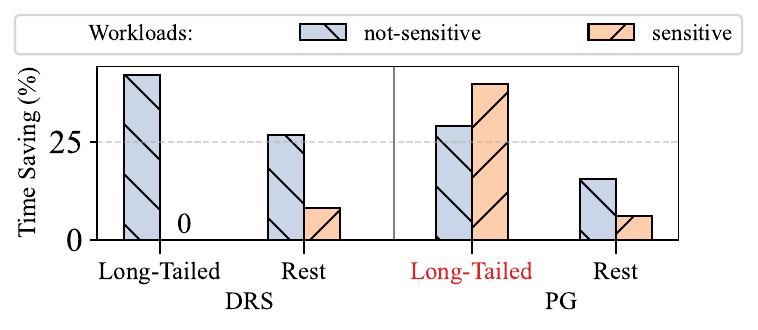}
    \vspace{-3mm}
    \caption{MV-driven time saving split by query latency quantile (Long-Tailed = top 10\% by original latency vs Rest = bottom 90\%) on PostgreSQL and Doris; \tpcds{} (balanced) vs \dsb{} (skewed) instances.}
    \label{fig:time_saving_by_quantile}
    \vspace{-3mm}
\end{figure}

\textbf{Long-tail queries domination.} Another discovery from Table \ref{tab:skewed_distribution_ts} is that sensitive workload always performs better than not-sensitive one in PostgreSQL, and this still holds even under instance mismatched scenario. It is an engine optimizer level heterogeneity as in Doris the situation is reversed. Our further breakdown shows such difference is determined by the key minorities. Splitting the workload by original latency (Long-Tailed = top 10\% by original latency; Rest = bottom 90\%), Figure~\ref{fig:time_saving_by_quantile} shows that (Long-Tailed queries, PostgreSQL) is the only group where it is the sensitive workload that enjoys higher performance, which in turn pulls up the overall performance. Long-tailed queries are therefore the key axis of variation under data-distribution drift whose plans are most exposed to skew-induced cardinality-estimation errors.

\textbf{Case study: how MVs offset skew on a long-tailed sensitive query.} To explain the mechanism we profile a single query drawn from the long-tailed sensitive workload on \pg{} (Table~\ref{tab:skewed_distribution_case_study}). The original query is 13.75$\times$ (72.48/5.27) slower on \dsb{} than on \tpcds{} and processes 10.73$\times$ (118/11) more buffer blocks. After MV rewriting, both latency and buffer-block counts return to the \tpcds{}-level baseline. Plan inspection reveals the mechanism: on \tpcds{}, \pg{}'s optimizer estimates that the index scan on \texttt{store\_sales} will return 29 rows and chooses a nested-loop join (manageable: 20\,448 actual loops). On \dsb{}, the same plan combined with skewed cardinality (estimated 1 row, actual 34\,M loops) explodes into 101\,M buffer blocks. The MV-rewritten plan switches to a hash join with a sequential scan over the materialized view, saving 68.68\,s and 112\,M buffer blocks on this single query.

{
\begin{table}[!htp]\centering\small
\caption{Per-plan profile for the long-tailed sensitive case-study query on \pg{} before and after MV rewriting, on \tpcds{} (balanced) and \dsb{} (skewed) instances. ``Key Node'' is the dominant join operator on \texttt{store\_sales}.}\label{tab:skewed_distribution_case_study}
\scriptsize
\begin{tabular}{p{2.5cm}rrrrr}\toprule
&\multicolumn{2}{c}{Original Query} &\multicolumn{2}{c}{Rewritten Query} \\\cmidrule{2-5}
Metrics &\tpcds &\dsb &\tpcds &\dsb \\\midrule
Latency (s) &5.27 &\textcolor{red}{72.48} &3.92 &3.80 \\
Buffer Blocks (M) &11 &\textcolor{red}{118} &6 &6 \\
Key Node &\multicolumn{2}{c}{Nested Loop + Index Scan} &\multicolumn{2}{c}{Hash Join + Seq Scan} \\
Estimated Rows &29 &\textcolor{red}{1} &147\,104 &53\,654 \\
Actual Loops &20\,448 &\textcolor{red}{34\,M} &8 &8 \\
Node Buffer Blocks (M) &6 &\textcolor{red}{101} &0.9 &0.9 \\
\bottomrule
\end{tabular}
\vspace{-3mm}
\end{table}
}

\textbf{Rewritability under data-distribution drift varies sharply by engine.} Under instance mismatched scenario, the per-engine rewrite-success rate also diverges dramatically when the same \ecse-enumerated view set is shifted from not-sensitive to sensitive workload. Table~\ref{tab:skewed_distribution_rewritability} summarizes the rewrite-success counts and percentages.

{
\begin{table}[!htp]\centering\small
\caption{Rewrite-success rate (\%) and rewritten pair count under \ecse{} enumeration with each engine's native rewriter, comparing not-sensitive and sensitive workloads. \doris{}'s rewritability collapses by 71\,pp under skew, while \starrocks{}'s rises by 31\,pp.}
\label{tab:skewed_distribution_rewritability}
\scriptsize
\begin{tabular}{lrrrr}\toprule
\textbf{Workload} & \textbf{\doris} & \textbf{\hive} & \textbf{\redshift} & \textbf{\starrocks} \\\midrule
not-sensitive  & 4298 (99\%) & 3890 (89\%) & 3994 (92\%) & 1119 (26\%) \\
sensitive   & 1113 (28\%) & 3562 (91\%) & 2736 (70\%) & 2249 (57\%) \\
\bottomrule
\end{tabular}
\vspace{-3mm}
\end{table}
}

Three patterns are visible: (i) \doris{}'s rewritability collapses from 99\% to 28\% --- its strict plan-pattern matching fails when skew-induced cardinality changes push the optimizer toward alternative join orders that the rewriter no longer recognizes. (ii) \starrocks{} \emph{increases} from 26\% to 57\% --- skew may force simpler join plans whose narrower shape falls within \starrocks{}'s stricter rewriter window. (iii) \hive{} stays nearly flat (89\%$\to$91\%) and \redshift{} drops moderately (92\%$\to$70\%), confirming that the rewriter's pattern-matching flexibility --- not just the underlying data --- mediates the data-distribution-drift effect.

\subsubsection{Hardware and Memory Pressure}
\label{sec:robustness-hw}
To stress the pipeline under resource pressure, we decouple the environment used to collect training data, make recommendation decisions and rewrite from the environment used to benchmark the performance of recommended MVs. Each environment has two options: \emph{constrained} (16 vcpus and 32GB) versus \emph{abundant} (64 vcpus and 128GB).

\textbf{Selector behavior under memory pressure.} Table~\ref{tab:memory_pressure_performance} reports workload time saving in the cross-hardware transfer study. "Native" scenario means the machine used for MV recommendation is the same as the evaluation environment while "Cross" is not. Firstly, the cross-hardware settings achieve workload time savings comparable to the non-cross settings, with \uniview-based pipelines even \emph{benefits} from memory pressure (cross-native difference 7.65\% and 8.38\%). In other words, MVs recommended on the abundant machine remain highly effective when evaluated on the constrained machine, and vice versa. Our breakdown analysis (Table \ref{tab:ts_breakdown_by_mv_source}) further shows in either evaluation environments, it is the MVs recommended from both environments that contribute the most time saving (34.91\% and 36.86\%)---valuable MVs can always be discovered no matter which hardware is adopted. All theses strongly indicate that MV recommendation exhibits a considerable degree of transferability across hardware environments.

{
\begin{table}[!htp]\centering\small
\caption{Time Saving in Cross Hardware Transfer Study (\%)}\label{tab:memory_pressure_performance}
\scriptsize
\begin{tabular}{p{1cm}p{1.2cm}p{1.6cm}p{0.7cm}p{0.7cm}p{0.7cm}r}\toprule
& & &\multicolumn{3}{c}{\textbf{MV Recommended By}} \\\cmidrule{4-6}
Evaluation env. &Enumerator &Recommender &Native env. &Cross env. &Cross-Native \\\midrule
abundant &ECSE &BigSubs &49.78 &49.11 &-0.67 \\
abundant &UniView &BigSubs &61.60 &61.32 &-0.28 \\
abundant &ECSE &GnnMV &53.48 &53.89 &0.41 \\
abundant &UniView &GnnMV &60.18 &59.71 &-0.47 \\\midrule
constrained &ECSE &BigSubs &56.91 &52.17 &-4.74 \\
constrained &UniView &BigSubs &63.34 &70.99 &7.65 \\
constrained &UniView &GnnMV &62.85 &71.23 &8.38 \\
constrained &ECSE &GnnMV &57.83 &58.66 &0.83 \\
\bottomrule
\end{tabular}
\end{table}

\begin{table}[!htp]\centering\small
\caption{Time Saving (\%) Breakdown By the Source of Recommended MV. Enumerator=\ecse, Recommender=\gnn.}\label{tab:ts_breakdown_by_mv_source}
\scriptsize
\begin{tabular}{lrrr}\toprule
&\multicolumn{2}{c}{Evaluation env.} \\\cmidrule{2-3}
MV recommended by & abundant&constrained \\\midrule
both env. &34.91 &36.86 \\
abundant only &18.57 &21.80 \\
constrained only &18.98 &20.97 \\
\bottomrule
\end{tabular}
\end{table}
}

\textbf{Rewriting behavior under memory pressure.} We also re-examine rewrite success rate (defined same as the one in Table \ref{tab:exp-rewrite-success}) of each rewriter and conclude the following findings (Table \ref{tab:rewrite_success_rate_hardware_mix}): (a). Regardless of rewriting environment, \hive~ is the only rewriter consistently enjoys high level success rate across both enumerator, while \redshift~ and \starrocks~ only performs well in \ecse~ and \uniview~ respectively. (b). Under memory pressure, only \hive~ and \redshift~ do not experience degradation and \hive~ even increases 11\% on \ecse. 

{
\begin{table}[!htp]\centering\small
\caption{Rewrite Success Rate (\%) Under Different Hardwares.}\label{tab:rewrite_success_rate_hardware_mix}
\scriptsize
\begin{tabular}{p{1.3cm}p{0.8cm}rrrrrr}\toprule
Rewriting Env. &Enumerator &DRS &HIV &Sys-A &STR \\\midrule
abundant &ECSE &15\% &66\% &70\% &35\% \\
constrained &ECSE  &10\% &77\% &69\% &46\% \\
abundant &UniView &2\% &85\% &6\% &73\% \\
constrained &UniView &1\% &85\% &6\% &66\% \\
\bottomrule
\end{tabular}
\end{table}
}

\begin{tcolorbox}[colback=gray!10,colframe=gray!40,boxrule=0.5pt,arc=2pt,left=5pt,right=5pt,top=5pt,bottom=5pt]
\noindent\textbf{Robustness Takeaways.}

\textbf{(RQ1)} \ecse{} is more robust to workload drift than \uniview{}: \ecse{}'s drift-induced shrink factors (1.38 and 0.73 on \job{} and \scale{} respectively) stay near 1, while \uniview{}'s reach 3.18 and $+\infty$ (\textsection\ref{sec:robustness-drift}). Cross-workload training can even outperform in-distribution training when the training workload exposes a richer join structure (\ecse{} on \scale-rewrite, shrink factor 0.73).

\textbf{(RQ1/RQ3)} MV rewriting can offset, not merely suffer from, data-distribution skew. On the long-tailed sensitive workload on \pg{}, the original query is 13.75$\times$ slower on \dsb{} than \tpcds{}, but MV rewriting returns latency to the \tpcds{}-baseline by switching nested-loop+index-scan to hash-join+seq-scan, saving 68.56\,s and 112\,M buffer blocks on a single query (\textsection\ref{sec:robustness-data}).

\textbf{(RQ2)} It is safe to apply recommended MVs on cross hardware evaluation as recommenders show robust transferability on memory pressure scenario. However, rewriter should be picked up with caution as certain cost-sensitive optimizers (\doris~ and \starrocks) are expected to flucturate on the amount of rewritten queries. Overall, \hive~ is still the best recommendation as it is not only robust to hardware but also to different enumerators. (\textsection\ref{sec:robustness-hw}).
\end{tcolorbox}

\section{Diagnosing Rewrite Gaps Between Hive and Doris}
\label{sec:case-rewriter}
In section \ref{sec:exp-rewriter}, we showed that Hive's SQL-transparent rewriter often outperforms other engines' native optimizer-level rewriters. In this case study, we focus on the \stats{} workload with \ecse{} candidates and compare \doris{} against \hive{}. We analyze query--view pairs that are successfully rewritten by \hive{} but not by \doris{}. We choose this setting for two reasons: (i) (\stats{}, \ecse{}) yields the largest number of candidate rewrite pairs (3,256), providing a large and diverse sample; and (ii) \doris{} exposes diagnostic information in its optimizer output, which helps attribute rewrite failures.

{
\begin{table}[t]\centering
\caption{$(q, v)$ pairs rewritten by \hive~ but not \doris.}\label{table:rewriting_failure_pairs_with_doris_reason_and_hive_advantage}
\vspace{-2mm}
\scriptsize
\begin{tabular}{p{3.4cm}p{1cm}p{3cm}}\toprule
\textbf{Failure reason from Doris} &\textbf{\# Pairs} &\textbf{Advantages of Hive} \\\midrule
Rewrite success but not selected &7 &- \\\midrule
View struct info is invalid &2408 &$\mathsf{A1}$: Join reordering and $\mathsf{A2}$: Predicate pushdown \\\midrule
View struct info is invalid, Predicate compensate fail &38 &$\mathsf{A3}$: Precise predicate compensation \\
\bottomrule
\end{tabular}
\end{table}
}

\noindent\textbf{High-level failure breakdown.}
Table~\ref{table:rewriting_failure_pairs_with_doris_reason_and_hive_advantage} reports three categories of  \hive{}-only pairs and the number of affected pairs. 
We identify 7 pairs that \doris{} can rewrite but does not choose in the final plan, suggesting a \emph{rewrite-found-but-pruned} behavior (may be due to cost-based optimization). Since these account for only 0.3\% of \hive{}-only cases and cost models are highly engine-specific, we do not treat cost-based pruning as a primary source of the observed gap.

\noindent\textbf{Diagnosing structural non-rewritability in Doris.}
The remaining gap is dominated by structural rewrite failures in \doris{}. While \doris{} labels most failures with coarse reasons (e.g., ``View struct info is invalid''), only 38 instances (1.55\%) provide a more specific message (``Predicate compensate fail''); the remaining 2,408 instances lack an actionable explanation. We therefore inspect the query and view structures and summarize the dominant failure patterns into two advantages of \hive{} (A1 and A2).

\noindent\textbf{Drilling down via high-impact views.}
We group the 2,408 failures by view and find that a small number of views accounts for a large fraction of failures: the top 5 (out of 53) cover 56\% of all failures. We select the most frequent view $v_1$ with a simple join structure for detailed inspection because (i) it alone accounts for 16\% of failures (402 pairs), (ii) it also has 13 successful rewrites in \doris{}, enabling controlled comparison under the same view.

\noindent\textsf{A1} \textbf{Flexibility in join matching.}
Queries $q_1$ and $q_2$ in Figure~\ref{figure:rewritability_hive_vs_doris_example_queries_in_sql} differ only in that $q_2$ joins one additional relation (\textit{posts}). Both queries contain the join \textit{users} $\Join$ \textit{badges}, which is exactly the view $v_1$, and both queries require only attributes covered by $v_1$ for this join. Thus, replacing \textit{users} $\Join$ \textit{badges} with a scan of $v_1$ should be valid in both cases. Empirically, however, $q_1$ is rewritten by both \hive{} and \doris{}, while $q_2$ is rewritten only by \hive{}. Inspecting the 13 \doris{} successes for $v_1$, we find that \doris{} rewrites only when the surrounding join graph belongs to two specific relation-set patterns; queries with other relation sets (including $q_2$) are rewritten only by \hive{}. This suggests that \doris{} is less flexible in substituting a matched view into larger joins (or requires stricter join-order/pattern conditions).

\noindent$\mathsf{A2}$ \textbf{Predicate pushdown limitations.}
$q_1$ and $q_3$ share the same join, but $q_3$ has additional selection predicates on \textit{users}. Since $v_1$ has no predicates, these extra filters are safe to apply on top of a scan of $v_1$. However, \hive{} rewrites $q_3$ using $v_1$, while \doris{} does not, indicating a more restrictive predicate-compatibility implementation.

\begin{figure}
\begin{lstlisting}[basicstyle=\scriptsize\ttfamily,numbers=left,language=SQL]
v1: SELECT * FROM users AS u JOIN badges AS b ON u.Id = b.UserId;
\end{lstlisting}
\begin{lstlisting}[basicstyle=\scriptsize\ttfamily,numbers=left,language=SQL]
v2: SELECT * FROM posts AS p JOIN votes AS v ON 
p.OwnerUserId = v.UserId JOIN users AS u ON p.OwnerUserId = u.Id 
WHERE p.CommentCount >= 27; -- more general filtering value
\end{lstlisting}
\vspace{-3mm}
\begin{lstlisting}[basicstyle=\scriptsize\ttfamily,numbers=left,language=SQL]
q1: SELECT COUNT(*) 
FROM comments as c, postHistory as ph, badges as b, users as u 
WHERE u.Id = c.UserId AND u.Id = ph.UserId AND u.Id = b.UserId AND c.Score = 18 AND c.CreationDate >= '2009-04-23 07:14:48'::timestamp AND c.CreationDate <= '2011-09-05 18:06:40'::timestamp AND ph.PostHistoryTypeId = 1 AND ph.CreationDate >= '2012-03-25 03:12:39'::timestamp AND ph.CreationDate <= '2012-12-24 19:24:20'::timestamp AND b.Date <= '2012-01-08 08:42:47'::timestamp;
\end{lstlisting}
\vspace{-3mm}
\begin{lstlisting}[basicstyle=\scriptsize\ttfamily,numbers=left,language=SQL]
-- q2 involves one additional relation posts compared to q1
q2: SELECT COUNT(*) 
FROM comments as c, posts as p, postHistory as ph, badges as b, users as u
WHERE u.Id = ph.UserId AND u.Id = b.UserId AND u.Id = p.OwnerUserId AND u.Id = c.UserId AND c.Score=12 AND p.PostTypeId=1 AND p.ViewCount>=46 AND p.ViewCount<=33342 AND p.FavoriteCount=6 AND p.CreationDate<='2013-09-24 04:55:16'::timestamp;
\end{lstlisting}
\vspace{-3mm}
\begin{lstlisting}[basicstyle=\scriptsize\ttfamily,numbers=left,language=SQL]
-- q3 involves more selection predicates on users compared to q1
q3: SELECT COUNT(*) 
FROM comments as c, postHistory as ph, badges as b, users as u 
WHERE u.Id = b.UserId AND u.Id = ph.UserId AND u.Id = c.UserId AND c.CreationDate<='2013-04-09 10:50:59'::timestamp AND b.Date>='2013-10-27 08:37:08'::timestamp 
AND u.Reputation>=23996 AND u.Reputation<=66025 AND u.DownVotes>=20 AND u.DownVotes<=100; 
\end{lstlisting}
    \vspace{-4mm}
    \caption{Queries and views for which \hive{} and \doris{} exhibit different rewriting behavior.}
    \label{figure:rewritability_hive_vs_doris_example_queries_in_sql}
    \vspace{-4mm}
\end{figure}
\section{Risky Rewrite Features} \label{sec:risky_rewrite}

In Table \ref{tab:exp-rewrite-success} \textsection \ref{sec:exp-rewriter}, we observe \starrocks~ and \celerdata~ achieve best success rewrite ratio on \scale, unlike their performance on other workloads. However, the net time saving result on their successfully rewritten queries shows such high rewrite yields to a catastrophic performance regression (Table \ref{tab:risky_rewrite_time_saving}). Specifically, \starrocks~ reports a catastrophic net time saving of $-111.83\%$, effectively more than doubling the query execution time. Similarly, \celerdata~ exhibits a net time saving of $-50.36\%$. An inspection of the generated physical plans reveals that the rewriter introduces an approach we term \textit{Inverse Query Reconstruction} (illustrated in Figure \ref{fig:risky_rewrite_time_saving_query}). When the available MV contains only a strict subset of the target relation (\texttt{title}), the rewriter attempts to compensate for the missing tuples. It achieves this by synthesis: generating a complex \texttt{UNION ALL} subquery that pairs the scan of the MV with a fallback scan over the original base table title, guarded by a disjunctive chain of inverse predicates (\texttt{kind\_id <= 1 OR production\_year <= 2013 OR production\_year IS NULL}).

In fact, such rewritings are certainly highly risky and prone to suffer from optimization barrier: Encapsulating the base relation within a structural \texttt{UNION ALL} subquery introduces an optimization barrier. This complex set operations pipeline limits the CBO capacity to perform join reordering and blocks crucial downstream predicate pushdowns from outer relational operators into the base table scan. Additionally, reconstructing the relation dynamically introduces substantial materialization overhead. The engine must maintain extra pipeline buffers to merge the data streams from both the MV scan and the high-volume base table scan before routing the unified stream to subsequent join stages, massively driving up memory consumption and network shuffle traffic.

\emph{How to detect such risky rewrites?} To prevent such malicious rewrites in production frameworks, we propose users can develop a static structural checker. If a candidate plan introduces a \texttt{UNION ALL} structure specifically for base-table data compensation using inverse predicates, and those predicates feature non-trivial disjunctions (OR or NOT IN), the rewrite should be eagerly flagged as high-risk.

{
\footnotesize
\begin{table}\centering
\caption{Net Time Saving (\%) of Queries Extra Rewritten by \starrocks~ and \celerdata}\label{tab:risky_rewrite_time_saving}
\begin{tabular}{lp{2cm}p{2cm}p{2cm}r}\toprule
\textbf{Rewriter} &\textbf{Extra Rewritten Pair Count} &\textbf{Positive Speedup Query Ratio (\%)} &\textbf{Net Time Saving (\%)} \\\midrule
STR &1176 &29.25 &-111.83 \\
COM-ii &1175 &9.19 &-50.36 \\
\bottomrule
\end{tabular}
\end{table}

\begin{figure}
    \centering
    \begin{lstlisting}[basicstyle=\tiny\ttfamily,numbers=left,language=SQL] 
    -- original query
    SELECT  COUNT(*)
    FROM title t, movie_companies mc, cast_info ci, movie_info mi, movie_keyword mk
    WHERE t.id = mc.movie_id AND t.id = ci.movie_id AND t.id = mi.movie_id AND t.id = mk.movie_id AND ci.person_id < 926305 AND ci.role_id < 10 AND mi.info_type_id < 7 AND mk.keyword_id > 1074;
    
    --recommended view (name: `mv`)
    SELECT ... 
    FROM title AS t
    WHERE t.kind_id > 1 AND t.production_year > 2013
    
    --rewritten query from STR / COM-ii using `mv`
    SELECT COUNT(*)
    FROM (
        SELECT id 
        FROM title
        WHERE kind_id <= 1 OR production_year <= 2013 OR production_year IS NULL
        UNION ALL  --KEY: inversely reconstruct the complete relation `title` partially using `mv`
        SELECT id FROM mv
    ) as t
    JOIN ......
    WHERE ci.person_id < 926305 AND ci.role_id  < 10 AND mi.info_type_id < 7 AND mk.keyword_id  > 1074;
    \end{lstlisting}
    \caption{Risky Rewrite Pattern Introduced by \starrocks~ and \celerdata}
    \label{fig:risky_rewrite_time_saving_query}
\end{figure}
}
\section{Filtering quality of view selectors. }\label{sec:precise_filtering}

Table \ref{tab:exp-selector-filter-quality} reveals a three-linked discoveries connecting view filtering to final savings. First, a high SVR does not imply high TS: \celerdata~ selects nearly all candidates on STATS (SVR $=$ 100\%) yet achieves only 0.78\% TS, while \gnnmv's more selective filtering (SVR $=$ 90.54\%) \textbf{precisely} generates 8.6× more rewrites (1,984 vs. 230) and reaches the workload-best 67.87\% TS (highlighted in green). Second, rewrite volume (number of successful rewrites) must be paired with per-rewrite quality (SQR): on \tpcds, \celerdata~ combines the highest rewrite count (469) with a perfect SQR of 100\%, delivering the best 57.72\% TS (highlighted in green), whereas \gnnmv's far fewer rewrites (67, SQR = 74.40\%) yield only 1.90\% TS despite comparable SVR. Third, the dominant selector is workload-dependent: \gnnmv~ leads on \job~ (55.71\%, green) and \stats~ (67.87\%, green) by prioritizing high-coverage views that enable broad and beneficial rewrites; \celerdata~ dominates on \tpcds~ (57.72\%, green) through compact, column-pruned views that unlock near-perfect rewrite applicability; and \bigsubs~ edges ahead on \scale~ (16.73\%, green) through a consistent balance of volume and speedup quality. Together, these results show that no single filtering metric — SVR, number of rewrites, or SQR in isolation — predicts end-to-end savings; only their joint alignment with the structural characteristics of the target workload determines which selector prevails.

\begin{table}[H]
\caption{Filtering quality of view selectors. SVR (selected-view ratio) SVR is the fraction of candidate views that are selected. \#rewrites is the number of queries that are successfully rewritten using selected views. SQR (speedup-query ratio) is the proportion of rewritten queries that yield a speedup. TS = Time saving ratio.}\label{tab:exp-selector-filter-quality}
\vspace{-3mm}
\scriptsize
\centering
\begin{tabular}{p{0.8cm}rC{1.2cm}C{1cm}C{1cm}C{1cm}}\toprule
\textbf{Workload} &\textbf{Selector} &\textbf{SVR (\%)} &\textbf{\#rewrites} &\textbf{SQR (\%)} &\textbf{TS (\%)} \\\midrule
\multirow{4}{*}{JOB} &\bigsubs{} &62.04 &\textbf{126} &34.04 &\textbf{39.10} \\
&\celerdata{} &\textbf{91.67} &19 &\textbf{100.00} &1.46 \\
&\gnnmv{} &\textbf{90.09} &\textbf{154} &\textbf{41.06} &\fancycellGreen{\textbf{55.71}} \\
&\uniview{} &86.82 &106 &40.42 &31.35 \\\hline
\multirow{4}{*}{SCALE} &\bigsubs{} &\textbf{94.03} &269 &\textbf{65.67} &\fancycellGreen{\textbf{16.73}} \\
&\celerdata{}&\textbf{90.54} &\textbf{281} &\textbf{93.95} &3.51 \\
&\gnnmv{} &87.82 &\textbf{278} &58.23 &\textbf{16.43} \\
&\uniview{} &82.99 &177 &59.25 &7.45 \\\hline
\multirow{4}{*}{STATS} &\bigsubs{} &99.07 &\textbf{1946} &49.13 &\textbf{62.23} \\
&\celerdata{} &\textbf{100.00} &230 &\textbf{96.96} &0.78 \\
&\gnnmv{} &95.54 &\textbf{1984} &\textbf{56.91} &\fancycellGreen{\textbf{67.87}} \\
&\uniview{} &\textbf{100.00} &554 &46.34 &27.68 \\\hline
\multirow{4}{*}{TPCDS} &\bigsubs{} &97.67 &\textbf{101} &65.70 &\textbf{4.99} \\
&\celerdata{} &\textbf{100.00} &\textbf{469} &\textbf{100.00} &\fancycellGreen{\textbf{57.72}} \\
&\gnnmv{} &\textbf{100.00} &67 &\textbf{74.40} &1.90 \\
&\uniview{} &94.44 &91 &32.14 &0.41 \\
\bottomrule
\end{tabular}
\end{table}

\section{Supplementary Material for Case 1: Comparing Non-commercial vs.
Sys-B Enumerator Across Workloads}

\begin{figure}[H]
    \centering
    \begin{lstlisting}[basicstyle=\tiny\ttfamily,numbers=left,language=SQL]
    -- original query
    SELECT COUNT(*) 
    FROM title t,movie_companies mc,movie_info mi,movie_info_idx mi_idx,movie_keyword mk 
    WHERE t.id=mc.movie_id AND t.id=mi.movie_id AND t.id=mi_idx.movie_id AND t.id=mk.movie_id AND t.production_year<2011 AND mc.company_id>1060 AND mk.keyword_id<31445;
    
    --view enumerated by Sys-B (0.13MB)
    SELECT t.kind_id,t.production_year,mi_idx.info_type_id
    FROM movie_info_idx AS mi_idx JOIN title AS t ON mi_idx.movie_id = t.id
    GROUP BY production_year,kind_id,info_type_id
    
    --view from ECSE (180MB)
    SELECT * 
    FROM title AS t JOIN movie_info_idx AS mi_idx ON t.id = mi_idx.movie_id
    \end{lstlisting}
    \vspace{-4mm}
    \caption{Example of view definitions by \celerdata{} and \ecse{}.}
    \label{fig:case-sysb-enumerator}
\end{figure}
\end{document}